\documentclass[iop]{emulateapj}

\newcommand{\Pstar}{P{\text{*}}}
\usepackage{amsmath}

\begin{document}

\title{The Nature of Turbulence in the Outer Regions of Protoplanetary Disks}

\author{Daniel. A. Gole\altaffilmark{1,2} and Jacob B. Simon\altaffilmark{2,3}}

\begin{abstract}
We carry out a series of local, shearing box simulations of the outer regions of protoplanetary disks, where ambipolar diffusion is important due to low ionization levels, to better characterize the nature of turbulence and angular momentum transport in these disks. These simulations are divided into two groups, one with far ultraviolet (FUV) ionization included, and one without FUV. In both cases, we explore a large range in diffusivity values.  We find that in the simulations without FUV, the properties of the turbulence are similar to the unstratified simulations of \cite{bai11a}; for a given diffusivity, the MRI can still be present so long as the magnetic field is sufficiently weak.  Furthermore, the dynamics of the mid-plane in these simulations are primarily controlled by the MRI.  In the FUV simulations on the other hand, the MRI-active FUV layers transport strong toroidal magnetic flux to the mid-plane, which shuts off the MRI. Instead, angular momentum transport at the mid-plane is dominated by laminar magnetic fields, resulting in lower levels of turbulent Maxwell stress compared to the no-FUV simulations.  Finally, we perform a temporal correlation analysis on the FUV simulations, confirming our result that the dynamics in the mid-plane region is strongly controlled by the FUV-ionized layers. 
\end{abstract}

\altaffiltext{1}{JILA, University of Colorado and NIST, 440 UCB, Boulder, CO 80309, USA}
\altaffiltext{2}{Department of Astrophysical and Planetary Sciences, University of Colorado, Boulder, CO 80309, USA}
\altaffiltext{3}{Department of Space Studies, Southwest Research Institute, Boulder, CO 80302, USA}



\section{Introduction} 

One of the most basic and fundamental processes at work in accretion disks is the removal of angular momentum from the orbiting gas, which allows this gas to accrete onto the central object.  A commonly accepted way for this transport to occur is turbulence, which allows adjacent annuli to exchange angular momentum on a macroscopic level, creating an effective enhanced viscosity or  ``turbulent viscosity" \citep{shakura73}.  The leading candidate for sourcing this turbulence is the magnetorotational instability  \cite[MRI;][]{balbus91,balbus98}, which generates vigorous turbulence from an initially weak (compared to thermal pressure) and sufficiently well-coupled (to the disk gas) magnetic field.  

The degree of coupling between the magnetic field and the gas is crucial to the efficacy of the MRI in transporting angular momentum.  In fully ionized disks, such as those around black holes, this coupling is very strong.  However, the colder environments present in protoplanetary disks lead to low ionization levels; the result is substantially weakened gas-field coupling such that the MRI cannot operate at full capacity throughout the disk.  Early considerations of low ionization effects on the MRI led to a model, put forth by \cite{gammie96}, in which large Ohmic diffusivity within the very weakly ionized mid-plane quenches the MRI there, creating a so-called ``dead zone".  At larger heights from the mid-plane, however, the gas is ionized by cosmic-rays \citep{gammie96}, X-rays \citep{igea99}, and FUV photons \citep{perez-becker11b}.

More recently, it has been realized that two other low-ionization, ``non-ideal" magnetohydrodynamic (MHD) effects, namely, the Hall effect (which arises from ion-electron drift) and ambipolar diffusion (AD; which arises from ion-neutral drift), are also important in protoplanetary disks \citep{wardle99,sano02,fromang02,bai09,bai11a}. The relative strength of all three non-ideal MHD effects is dependent primarily on the density and ionization of the gas and thus the radius away from the central star \citep{kunz04,desch04}. Broadly speaking, the Ohmic term dominates at high densities, the Hall term dominates at intermediate densities, and the ambipolar term dominates at low densities; thus, moving away from the star, the dominant non-ideal term transitions from Ohmic diffusion to the Hall effect to ambipolar diffusion.

Considerable theoretical work has been carried out to study the role that both the Hall effect and ambipolar diffusion play in determining the gas dynamics in protoplanetary disks.  For ambipolar diffusion, in particular, analytic work has shown that the relevant control parameter (at least in the linear limit) for the growth of the MRI is the ratio of the rate of ion-neutral collisions to the orbital frequency $\Omega$, defined to be $Am$.  If $Am \lesssim 1$, AD damps the MRI, although unstable modes will always exists for a sufficiently weak field \citep{blaes94, kunz04}. This damping has been the topic of several numerical studies as well, seeking to understand the full, non-linear evolution of the MRI under these conditions.  \cite{hawley98} studied this problem treating the ions and neutrals as separate fluids coupled with a collisional drag term.  They establish limits on the relevance of AD, showing that the fluid behaves essentially the same as the ideal limit for collision frequencies on the order of $100\Omega$, while for collision frequencies $\lesssim 0.01 \Omega$ the neutral fluid is quiescent, unable to be affected by the motion of the ions.  \cite{bai11a} study the non-linear evolution of the ambiploar MRI in an unstratified shearing box using a single fluid method.  In contrast to the two-fluid approach, in the single fluid MHD  limit, the ion-neutral coupling is strong (which is shown to be appropriate for protoplanetary disks \citealt{bai11b})  the ion density is a fixed fraction of the neutral density instead of obeying it's own continuity equation, and the ion-density makes up a negligible fraction of the total mass-density.  \cite{bai11a} showed that for any given $Am$ value, one can always find a magnetic field strength at which the MRI is present in its full nonlinear state. For smaller $Am$, weaker magnetic field is required, which in turn leads to a lower level of turbulence. 

With these results in hand, \cite{bai11b, bai11c} use complex chemical networks in combination with ionization models for FUV, X-rays, and cosmic rays to calculate the ionization fraction and non-ideal MHD parameters in the disk as a function of $z$.  They then combine these calculations with the $\alpha-\beta_{min}-Am$ relationship from their unstratified simulations to predict the turbulence and stress in the disk as a function of $z$, and estimate the accretion rate.  Modeling of this nature implicitly assumes that {\it each layer of the disk operates more or less independently}, an assumption that remains untested. Indeed, several effects are relevant in suggesting that processes cannot be height-independent, including vertical turbulent diffusion, explicit diffusion of magnetic fields due to non-ideal effects, buoyancy, and advective transport due to large scale velocity modes that have been described analytically \citep{lubow93, ogilvie98} and have been seen in some shearing box experiments \citep{oishi09,gole16}.  The combination of these effects may play a significant role in determining the steady state structure of the outer regions of disks and may predict a significantly different accretion rate. 

The first such work to self-consistently model the AD-dominated outer disk with vertical gravity included was carried out by \cite{simon13a} and \cite{simon13b}. These authors carried out local and vertically stratified calculations, in which AD was quite strong near the mid-plane.  This work included ionization of upper disk layers by FUV photons \citep{perez-becker11b} leading to a scenario similar to the \cite{gammie96} model; active regions vertically sandwich a region of weak or no turbulence. Furthermore, these works determined that a net vertical field is required to predict high enough accretion rates to match the expected life-time of disks around T-Tauri stars (assuming that disk accretion is at least approximately in steady-state).  

The theoretical work done to-date has established a new understanding of turbulence in the outer regions of protoplanetary disks.  However, as alluded to above, the precise nature of this turbulence, and in particular, the role of vertical transport of energy (in any form) has not yet been addressed in detail.  It is a primary motivation of this paper to better understand turbulence in the AD-dominated outer regions of protoplanetary disks, and in particular to address whether or not each vertical layer can be treated (largely) independently. 

Additionally, beyond purely theoretical considerations, it has now become possible to put direct observational constraints on turbulence in the outer regions of disks with ALMA. Predictions of the strength of MRI-driven turbulence in an AD-dominated disk \citep{simon15} have been directly tested with observations of molecular line broadening \citep{flaherty15,flaherty17,flaherty18} as well as dust settling \citep{okuzumi16,pinte16}. These observations have (surprisingly) suggested that turbulence in these regions is quite weak.  These considerations motivate a more precise understanding of the physics of the outer disk regions for two primary reasons.  First, the presence of weak turbulence has only been established for a few systems; if disks with strong turbulence do exist, it behooves us to explore this turbulence in more detail.  Second, if indeed all disks show significantly weak turbulence, then we need to understand precisely why the MRI is prevented from operating.  

Thus, both theoretical and observational considerations strongly motivate us to deepen our understanding of turbulence in the outer regions of protoplanetary disks.  In this work, we carry out a series of numerical simulations to do just that.   

This paper is organized as follows: in section 2 we describe the numerical algorithm used in our simulations and explain the physics and initial conditions of the simulations, in section 3 we present the results from control simulations with no FUV layers, in section 4 we demonstrate the gas and magnetic field behaviour in simulations with FUV layers and contrast them with results from section 3, in section 5 we do a temporal correlation analysis to characterize the nature of the MRI and vertical magnetic flux transport in this situation, and in section 6 we summarize and discuss our primary results.


\section{Numerical Algorithm, Parameters, and Initial Conditions}

We carry out a series of numerical simulations with {\sc Athena}, which is a second-order accurate Godunov flux-conservative code that uses constrained transport \cite[CT;][]{evans88} to enforce the $\nabla \cdot \mathbf{B} = 0$ constraint, the
third-order in space piecewise parabolic method (PPM) of \cite{colella84} for spatial reconstruction, and the HLLD Riemann solver \citep{miyoshi05,mignone07} to calculate numerical fluxes.  A full description of the {\sc Athena} algorithm
along with results showing the code's performance on various test problems can be found in \cite{gardiner05}, \cite{gardiner08}, and \cite{stone08}.

We employ the shearing box approximation \citep{hawley95} to simulate a small, co-rotating patch of an accretion disk.  Taking the size of the shearing box to be small relative to the distance from the central object, we define Cartesian coordinates $(x, y, z)$ in terms of the cylindrical coordinates ($R,\phi,z^\prime$) such that $x=R-R_0$, $y=R_0\phi$, and $z = z^\prime$.  The box co-rotates about the central object with angular velocity $\Omega$, corresponding to the Keplerian angular velocity at $R_0$. 
The equations of MHD, with ambipolar diffusion as the only non-ideal MHD term, in the shearing box approximation are,
\begin{subequations}
\begin{equation}
\frac{\partial \rho}{\partial t} + \nabla \cdot (\rho \mathbf{v}) = 0, \\
\end{equation}
\begin{equation}
\frac{\partial (\rho \mathbf{v})}{\partial t} + \nabla \cdot (\rho \mathbf{v} \mathbf{v} - \mathbf{B} \mathbf{B} + \mathbf{\bar{\bar{I}}} \Pstar) =  \\ 2 q \rho \Omega^2 x \mathbf{\hat{i}} - \rho \Omega^2 z \mathbf{\hat{k}}  - 2 \Omega \mathbf{\hat{k}} \times \rho \mathbf{v}, \\
\end{equation}
\begin{equation}
\frac{\partial \mathbf{B}}{\partial t} - \nabla \times \big[\mathbf{v} \times \mathbf{B} - \frac{(\mathbf{J} \times \mathbf{B}) \times \mathbf{B}}{\gamma_c \rho_i \rho} \big] \ = 0. 
\label{eqInduction}
\end{equation}
\begin{equation}
\frac{\partial E}{\partial t} + \nabla \cdot \big[(E+\Pstar)\mathbf{v}-\mathbf{B}(\mathbf{B}\cdot\mathbf{v})\big] = 0
\end{equation}
\begin{equation}
\mathbf{J} = \nabla \times \mathbf{B}
\end{equation}
\begin{equation}
E = \frac{P}{\gamma-1} + \frac{\rho (\mathbf{v} \cdot \mathbf{v})}{2} + \frac{\mathbf{B} \cdot \mathbf{B}}{2}
\end{equation}
\end{subequations}
\noindent Here $\rho$ is the gas density, $\rho_i$ the ion density, $\gamma_c$ is the coefficient of momentum transfer in ion-neutral collisions, $\mathbf{B}$ is the magnetic field, $\mathbf{J}$ is the current density, $\Pstar$ is the total pressure (related to the magnetic field and the gas pressure, $P$, via $\Pstar = P + {\mathbf{B} \cdot \mathbf{B}}/{2}$), $\mathbf{\bar{\bar{I}}}$ is the identity tensor, $\mathbf{v}$ is the velocity, $E$ is the total energy, and $q$ is the shear parameter defined by $q=-d\ln{\Omega}/d\ln{R}$, which is $3/2$ in this case of a Keplerian disk. 

We use an adiabatic equation of state with a constant background temperature profile.  The disk is heated and cooled towards this profile with a characteristic cooling/heating time $\tau_{\text{cool}}$.  This adds the following term to the energy evolution equation:   
\begin{equation}
\frac{\partial E}{\partial t } = - \frac{\rho (T-T_0)}{\tau_{\text{cool}} (\gamma-1)}   \end{equation}
where $T\equiv P/\rho$ and $T_0$ is the constant background temperature.  We set $\tau_{\text{cool}}=0.1 \Omega^{-1}$.  This choice of equation of state and cooling prescription is part of an effort to move towards more realistic simulations of disks that account for both non-ideal MHD effects and adiabaticity, a regime that has not been well explored by previous studies. While not the main focus of this paper, we have opted to include this improved thermodynamics treatment as a step towards this goal of more realistic calculations. 


We use the standard boundary conditions appropriate for the shearing box approximation. At the azimuthal boundaries we exploit the symmetry of the disk and apply periodic boundary conditions. At the radial boundaries, we apply shearing periodic boundaries; quantities that are remapped from one radial boundary to the other are shifted in the azimuthal direction by the distance the boundaries have sheared relative to each other at the given time \citep{hawley95}. Furthermore, the energy and azimuthal velocity are appropriately adjusted to account for the difference in angular momentum and orbital energy \cite[see][]{hawley95}. Additionally, these radial shearing-periodic boundaries are incorporated in such a way as to maintain conservation of conserved variables \cite[see][]{stone10}. In the vertical direction, we employ the modified outflow boundary condition of \cite{simon13a}, which are standard outflow boundaries with the gas density extrapolated exponentially into the ghost zones; this method reduces the artificial build-up of toroidal magnetic flux near the vertical boundaries.  Such outflow boundaries can lead to significant mass loss, particularly in the case of strong vertical magnetic flux \citep{simon13b}.  In order to keep the box in a steady state, we re-normalize the mass of the box at every time step. Finally, we employ Crank-Nicholson differencing to conserve epicyclic energy to machine precision. A full discussion of these methods and {\sc Athena}'s shearing box algorithm can be found in \cite{stone10}.

Our standard choice of domain size is $L_x \times L_y \times L_z = 2H \times 4H \times 8H$, where $H$ is the isothermal scale height, defined to be
\begin{equation}
H = \sqrt{2} \frac{c_{s}}{\Omega}.     
\label{eq_H_definition}    
\end{equation}
where $c_s$ is calculated using $T_0$.\footnote{While our simulations are not isothermal, the cooling timescale is sufficiently short that departures away from isothermality are minimal; see below}   We also carry out simulations with $4H \times 8H \times 8H$ to test convergence with domain size (see table \ref{tabSims}).  Our simulations without an FUV layer will only extend $5H$ in the vertical direction, as the low density regions at large $|z|$ will impose a very strict constraint on the time-step from the strong diffusion in those regions. 

We initialize the box with net-magnetic-flux in the vertical direction such that the $\beta_{z,0}$, defined as 
\begin{equation}
\beta_{z,0} = \frac{2P_{\text{mid}}}{B_{z,0}^2},                 
\end{equation}
\noindent 
is $\beta_{z,0} = 10^4$. 
Note that a factor of $4\pi$ has been subsumed into the definition of the magnetic pressure, following the standard definition of units in {\sc Athena} \citep{stone08}.  In the case of a constant vertical field, the onset of the MRI creates a very strong channel mode that can create numerical problems \citep{miller00}.  To avoid this, the initial field has a spatial dependence in the $x$ direction:
\begin{equation}
B_z = B_{z,0}\left[1+\frac{1}{4}\sin (k_x*x)\right],                   
\end{equation}
where $k_x=2\pi/L_x$, and $L_x$ is the length of the box in the radial direction.     

In order to seed the MRI, we apply random perturbations to the initial gas pressure and velocity. Following the procedure used by \cite{hawley95}, the perturbations are uniformly distributed throughout the box and have zero mean.  Pressure perturbations are a maximum of $2.5 \%$ of the local pressure, and velocity perturbations are a maximum of $5\times 10^{-3}c_s$.

We initialize the disk is in hydrostatic equilibrium.  In an vertically isothermal disk, balancing gravity against the gas pressure gradient yields a Gaussian density profile.  While the disk is not isothermal in this case, it is heated/cooled towards an isothermal profile.  The heating/cooling time is shorter than the dynamical time, meaning that temperature perturbations will be restored to the isothermal temperature on a faster time-scale than the disk's density profile will respond on the scale of $H$.  For this reason, we will consider $H$ to be constant in our initialization and discussion.  We set $\rho_0 = 1$ as the initial density at the mid-plane.  A density floor of $10^{-4}$ is applied to keep the Alfv\'en speed from getting too high (and thus the time step too small) in the region far from the mid-plane.  The density floor also prevents $\beta$ from getting too low, which can cause numerical problems. We set the constant, background isothermal sound speed $c_s=1/\sqrt{2}$, and choose the angular velocity to be $\Omega = 1$, which gives $H = 1$.

\subsection{Resistivity Profile}
\label{profile}
For the FUV-ionized layers of the disk we use a resistivity model informed by the considerations of \cite{perez11} and adapted from other works \cite[e.g.,][]{simon13b}.  FUV radiation will penetrate some constant column density of the disk given by $\Sigma_{\text{FUV}}$.  We will call the point at which this column density occurs $z_{\text{FUV}}$.  As the box evolves, turbulence and magnetic pressure can be significant enough to change the density profile.  As a result, we need to re-calculate $z_{\text{FUV}}$ at each time-step.  This is done by integrating the horizontally averaged density profile from the top of the box towards the mid-plane until $\Sigma_{FUV}$ is reached.  The same calculation is made for the bottom half of the box separately.  

We will describe the strength of ambipolar diffusion in terms of the ambipolar Elsasser number, $Am$, defined as 
\begin{equation}
Am=\frac{\gamma \rho_i}{\Omega}.  
\end{equation}
$Am$ in the active layer ($|z|>z_{\text{FUV}}$) is generally $\gg$ 1.  We will use the prescription given in \cite{bai13a}:
\begin{equation}
Am = 3.37 \times 10^7 \bigg(\frac{f}{10^{-5}}\bigg) \bigg( \frac{\rho}{\rho_{\text{mid}}} \bigg) \bigg(\frac{r}{1 \text{AU}}\bigg)^{-5/4}.
\label{eqAmCorona}
\end{equation}
We take the ionization fraction $f$ to be $10^{-5}$ and place the box at $r=115$ AU in a minimum mass solar nebula model \citep{weidenschilling77, hayashi81}.

At the mid-plane the ionization fraction is generally controlled by X-rays and cosmic rays, which are able to penetrate farther into the disk.  \cite{bai11c} model this ionization using a complex chemical reaction network and show that $Am$ in the mid-plane region ($|z|<z_{\text{FUV}}$) is generally constant as a function of space and is on the order of 1.  Including small grains in this model decreases $Am$ and gives it a dependence on $\beta$.  We will not include this more complicated behaviour, but will consider a range of values for $Am_{\text{mid}}$ to account for this uncertainty.   The ionization profile will be smoothed over a thin region to prevent potential numerical problems, following the same procedure as \cite{simon13a}.  Our choice of $\Sigma_{\text{FUV}}$ is $0.1 {\rm g/cm^2}$.

\subsection{Resolution Considerations}
\begin{figure}[h!]
\centering
\includegraphics[width=0.9\columnwidth]{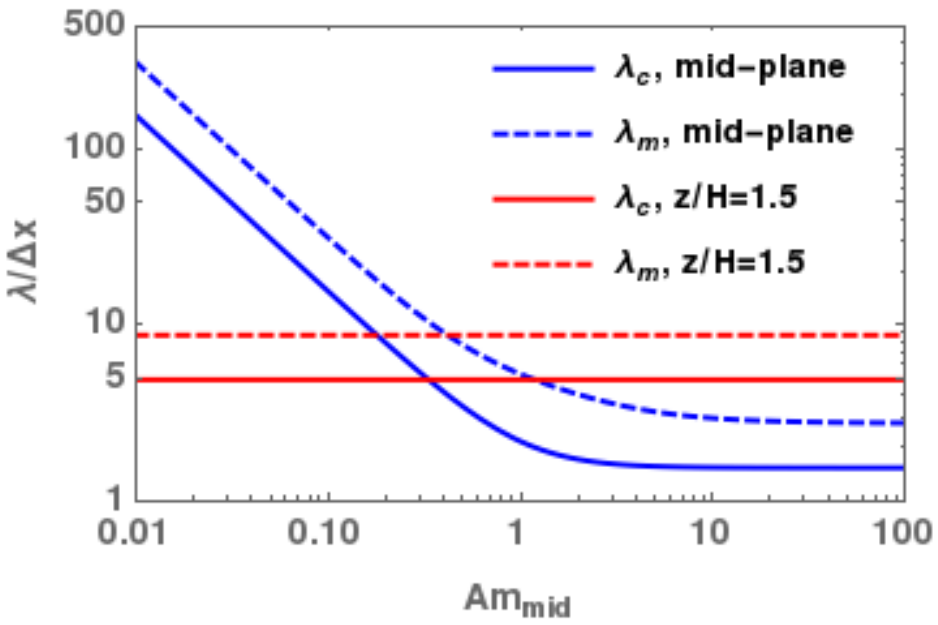}
\caption{The critical and most unstable MRI wavelengths at the mid-plane and at $z/H=1.5$) in terms of our fiducial cell size $\delta x = H/30$.  $\beta_0=10^4$ at the mid-plane, which corresponds to $\beta \sim 10^3$ at $z/H=1.5$ due to the decrease in gas pressure.  Am at $z/H=1.5$ is independent of Am at the mid-plane based on our prescription (see Section~\ref{profile}); we include the critical and most unstable MRI wavelengths, which do not depend on Am$_{\rm mid}$, at this location for reference.}
\label{figMriWavelength}
\end{figure}

Our fiducial choice of resolution for these simulations is 30 grid cells per $H$.  This choice is based on the size-scales on which the MRI operates in the linear limit.  This is typically characterized by the critical wavelength $\lambda_c$ and the most unstable wavelength $\lambda_m$.  All size-scales larger than $\lambda_c$ are unstable, and $\lambda_m$ corresponds to the mode withf the largest growth-rate.  For ideal MHD, $\lambda_m/H=9.18 \beta_0^{-1/2}$, while $\lambda_c$ is roughly half that \cite[see][]{hawley95}.  However, ambipolar diffusion increases both of these wavelengths significantly when $Am \lesssim 10$ \citep{wardle99, bai11a}.  We will consider a variety of $Am$ parameters for the mid-plane region, spanning $10^{-2}$ to $10^2$, while $Am$ will always follow equation \ref{eqAmCorona} for $|z|>z_{\text{FUV}}$. Figure \ref{figMriWavelength} shows the most unstable and critical wavelengths at the mid-plane and at $z/H=1.5$ for our choice of $\beta_0=10^4$.  The MRI is reasonably well resolved for $z>z_{\text{FUV}}$, with 9 grid cells per $\lambda_m$ at $z/H=1.5$.  However, at the mid-plane our resolution only yields 3 grid cells per $\lambda_m$ for $Am~\gtrsim~1$ .  While this is not ideal, previous convergence studies have shown that non-linear state of the MRI is not greatly changed even if, in the linear limit, it is under-resolved \citep{simon13b}.  This is potentially due to two effects.  First, even in the linear limit there are larger unstable modes than $\lambda_m$ that will be better resolved.  Second, as the MRI evolves into the non-linear regime the field will generally get stronger, increasing the most unstable size-scale such that it is better resolved.  

We should also note that at very high diffusivities, $\lambda_c$ can exceed the size of the ambipolar damping zone, which is $\approx 2H$, for our fiducial choice of parameters.  For example, at $Am=0.01$, $\lambda_c \sim 4H$ for $\beta_0=10^4$.  If no MRI modes fit within the damping zone, then this is another way in which ambipolar diffusion can effectively quench the MRI, in addition to that described in more detail below.  However, it is also worth mentioning that this argument only applies to vertical MRI modes and not toroidal or radial modes, which may still persist within the damping region.  We will also conduct simulations with no FUV layers, for which the vertical extent of the box is $5H$. In these simulations, our most diffusive case will be $Am=0.032$, and at this diffusivity both $\lambda_c$ and $\lambda_m$ fit within the vertical extent of the domain.

\subsection{List of Simulations and Varied Parameters}
Our main control parameter for this investigation will be $Am_{\text{mid}}$.  In order to rule out other effects and help confirm our results, we will run a few simulations at higher resolution, and a few with a larger domain in the horizontal directions.  In addition to the simulations with the FUV layer, we will perform simulations with constant $Am$ across the domain as a comparison.  In table \ref{tabSims} we list all of our simulations.  Our labeling convention is as follows: The first letter is S or L for small or large box size.  The second letter is L or H for low or high resolution.  The next letter will be F or N for simulations with FUV or no FUV respectively.  The number following Am is the $Am$ parameter at the mid-plane.
\begin{deluxetable}{l c c c c c}
\tablecaption{Simulation Parameters}
\tablewidth{0.5\textwidth}
\startdata
\hline \hline
Label & $Am_{\text{mid}}$ & FUV & $H/\Delta x$ & $(Lx, Ly, Lz)/H$ & Orbits \\
\hline
SLF\_Am0.01   & 0.01  & yes & 30 & 2,4,8  & 100    \\
SLF\_Am0.018  & 0.018 & yes & 30 & 2,4,8  & 50     \\
SLF\_Am0.032  & 0.032 & yes & 30 & 2,4,8  & 100    \\
SLN\_Am0.032  & 0.032 & no  & 30 & 2,4,8  & 75    \\
SLF\_Am0.056  & 0.056 & yes & 30 & 2,4,8  & 200    \\
SLF\_Am0.1    & 0.1   & yes & 30 & 2,4,8  & 400    \\
SLN\_Am0.1    & 0.1   & no  & 30 & 2,4,8  & 300    \\
LLF\_Am0.1    & 0.1   & yes & 30 & 4,8,8  & 100    \\
SHF\_Am0.1    & 0.1   & yes & 60 & 2,4,8  & 25     \\
SLF\_Am0.18   & 0.18  & yes & 30 & 2,4,8  & 400    \\
SLF\_Am0.32   & 0.32  & yes & 30 & 2,4,8  & 400    \\
SLN\_Am0.32   & 0.32  & no  & 30 & 2,4,8  & 300    \\
SLF\_Am0.56   & 0.56  & yes & 30 & 2,4,8  & 400    \\
SLF\_Am1      & 1.0   & yes & 30 & 2,4,8  & 400    \\
SLN\_Am1      & 1.0   & no  & 30 & 2,4,8  & 300    \\
SHF\_Am1      & 1.0   & yes & 60 & 2,4,8  & 200    \\
LLF\_Am1      & 1.0   & yes & 30 & 4,8,8  & 400    \\
SLF\_Am1.8    & 1.8   & yes & 30 & 2,4,8  & 400    \\
SLF\_Am3.2    & 3.2   & yes & 30 & 2,4,8  & 400    \\
SLN\_Am3.2    & 3.2   & no  & 30 & 2,4,8  & 300    \\
SLF\_Am5.6    & 5.6   & yes & 30 & 2,4,8  & 400    \\
SLF\_Am10     & 10    & yes & 30 & 2,4,8  & 100    \\
SLN\_Am10     & 10    & no  & 30 & 2,4,8  & 300     \\
SLF\_Am18     & 18    & yes & 30 & 2,4,8  & 100    \\
SLF\_Am32     & 32    & yes & 30 & 2,4,8  & 100    \\
SLF\_Am56     & 56    & yes & 30 & 2,4,8  & 100    \\
SLF\_Am100    & 100   & yes & 30 & 2,4,8  & 100    \\
\hline
\enddata
\label{tabSims}
\end{deluxetable}


\section{Stratified Ambipolar Simulations with no FUV}
\label{secNoFuv}

In this section we will consider a series of vertically stratified shearing boxes with no FUV-ionized layers (SLN\_Am0.032 through SLN\_Am10).  These will serve as a baseline for comparison with the FUV simulations.

\subsection{Expectations}
A simple way to consider the effect of resistive conditions on the MRI is to compare the MRI growth time-scale ($\sim \Omega^{-1}$) with the diffusion time-scale of the magnetic field.  If the diffusion time-scale is shorter than the growth time-scale, field will diffuse faster than the MRI can build it up and the MRI will not operate.  In the case of Ohmic diffusion, this is relatively straightforward because the Ohmic diffusion coefficient is not a function of the field strength.  So it is possible to say whether the MRI will operate or not based simply on the Elsasser number (which is itself essentially just a ratio of these time-scales).  However, ambipolar diffusion is more involved.  The ambipolar term in the induction equations expands into many terms (see equation \ref{eqInduction}).  Just for the sake of a heuristic argument, let's just consider the term that looks the most like simple diffusion, where the time derivative of a component is proportional to a double spatial derivative of the same component:
\begin{equation}
\bigg(\frac{\partial \mathbf{B}}{\partial t}\bigg)_{\text{AD}} = \frac{\mathbf{B}^2}{\Omega \rho \hspace{0.3mm} Am} \cdot \nabla^2 \mathbf{B}.
\end{equation}
This consideration is particularly valid in a shearing box for the $B_y$ component, which is expected to dominate over the other field components due to the shear in the disk.  Therefore this term with $B_y^2$ as a pre-factor will dominate over any cross-terms.  Notice that in this diffusion approximation, the diffusion coefficient itself scales with $B^2$, so the diffusion time-scale for length scale L does as well:
\begin{equation}
t_{\text{AD}} = \frac{L^2 \Omega \rho \hspace{0.3mm} Am}{B^2} = \bigg(\frac{L}{H}\bigg)^2 \frac{\beta \hspace{0.3mm}Am \hspace{0.3mm}}{\Omega},
\label{eqTad}
\end{equation}
with the second equivalency being true for vertically isothermal disk with constant sound speed, a reasonable approximation for our disk with an isothermal equilibrium temperature and a short cooling/heating time.  Considering a disk with a relatively weak seed-field, the MRI will be able to operate initially while the field is weak because the diffusion is also weak.  However, as the MRI winds-up the field, a saturation field strength will be reached where the diffusion time-scale and growth time-scales are now equal and the MRI no longer operates.  Setting $t_{\text{AD}}=\Omega^{-1}$ in the above equation yields the simple relation that the saturation $\beta$ parameter will be $\propto Am^{-1}$.

The fully non-linear version of this problem was considered by \cite{bai11a} in a 3D, unstratified, fully ambipolar diffusion dominated shearing box.  They fit the results of their simulations to show that the minimum $\beta$ parameter that permits the MRI is given by
\begin{equation}
\beta_{\text{min}}(Am)=\bigg[ \bigg( \frac{50}{Am^{1.2}} \bigg)^2 + \bigg( \frac{8}{Am^{0.3}} + 1 \bigg)^2  \bigg]^{1/2}.
\label{eqMinBeta}
\end{equation}
The simple expectation is roughly recovered for the limit $Am \lesssim 1$, where $\beta_{\text{min}} \propto Am^{-1.2}$.

\subsection{Magnetic Field Strength}
We characterize the magnetic field strength in terms of the $\beta$ parameter corresponding to the total field:
\begin{equation}
\beta = \frac{2P}{B_x^2 + B_y^2 + B_z^2}. 
\end{equation}
When space-time averages of $\beta$ are calculated, the pressure and total field averages are taken first, and then $\langle \beta \rangle$ is calculated from the ratio, i.e. $\langle \beta \rangle=\langle 2P  \rangle / \langle B \rangle$.  For the MRI operating in a shearing box, we expect $B_y$ to be the dominant field component as radial field will be sheared into toroidal field by the shear velocity profile.  In this case, $B \simeq B_y$ and $\beta \simeq \beta_y$.  

\begin{figure}[h!]
\centering
\includegraphics[width=0.9\columnwidth]{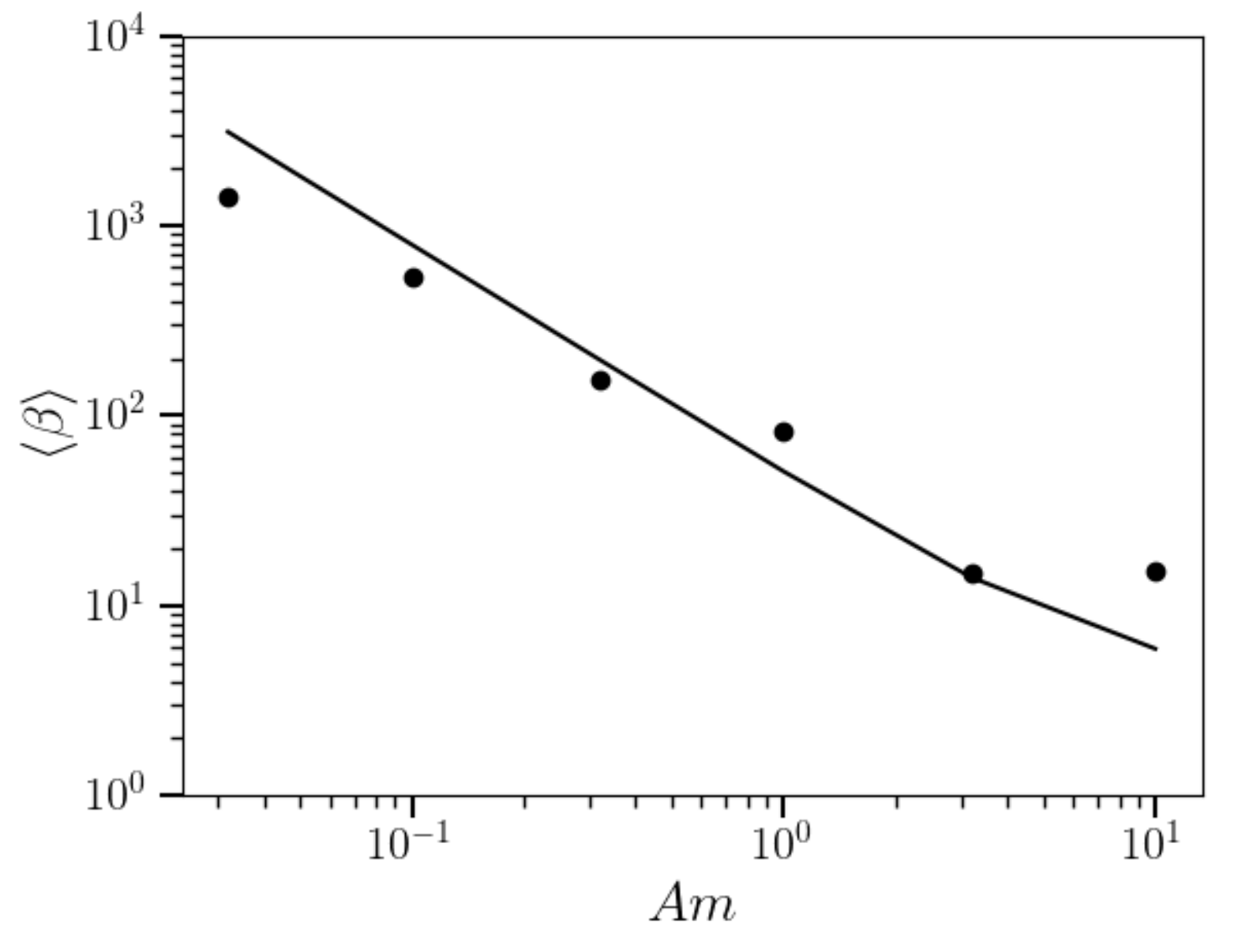}
\caption{Temporally and spatially averaged $\beta$ within $0.5H$ of the mid-plane.  The black line shows $\beta_{\text{min}}$ \cite[see][]{bai11a}.  $\beta_y$ generally follows $\beta_{\text{min}}$. }
\label{figBetaVsAmNFUV}
\end{figure}

In figure \ref{figBetaVsAmNFUV} we show the space-time averaged $\beta$ parameter in the mid-plane region of our stratified disk simulations.  We find that the average field strength in the saturated state closely follows the $\beta_{\text{min}}-Am$ relationship described by \cite{bai11a} for their unstratified simulations.  

\begin{figure}[h!]
\centering
\includegraphics[width=0.9\columnwidth]{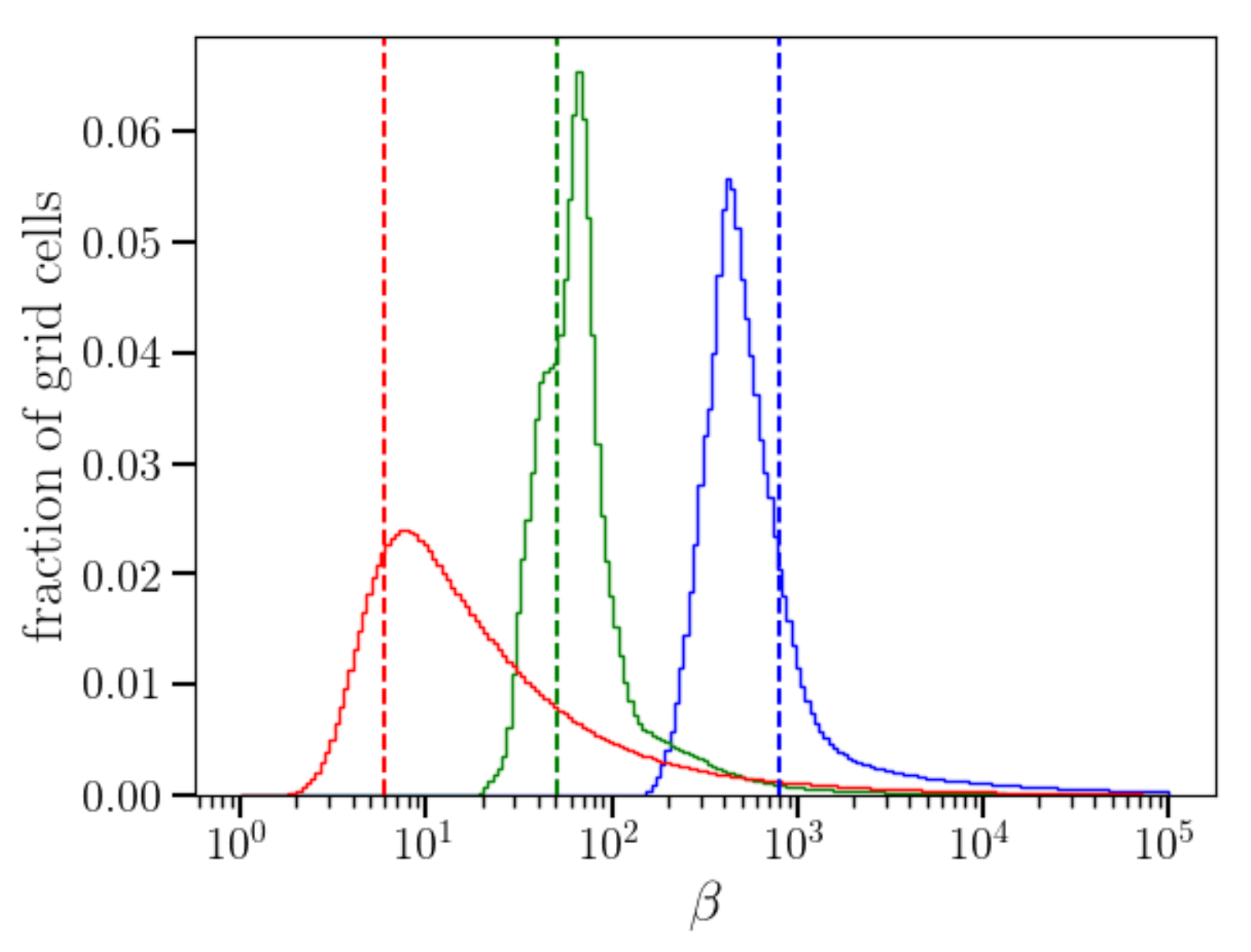}
\caption{$\beta$ distributions for simulations with $Am=0.1, 1, 10$ in blue, green, and red respectively.  The dashed vertical lines corresponds to $\beta_{\text{min}}$ for each simulation.  In general, $\beta_{\text{min}}$ gives a reasonably good description of the typical field strength.  }
\label{figBetaDistNFUV}
\end{figure}

We need to be cautious in using the average field strength, as it is possible that there exist regions of very strong field separated from weak field regions within the mid-plane (as shown by \citealt{simon17} in similar calculations); such non-uniform structure can lead to an average $\beta$ value that is not representative of the typical field strength found at the mid-plane.  It would thus be possible to sustain some turbulence sourced from small patches of weak-field ($\beta > \beta_{\text{min}}$) even if on average $\beta<\beta_{min}$.  We calculate the distribution of mid-plane field strength, integrated over time and for $|z|<0.5H$.  Examples of these distributions are shown in figure \ref{figBetaDistNFUV}.  Especially for the stronger diffusion cases, there are often regions where $\beta<\beta_{min}$, but the field strength does not exceed this limit drastically and $\beta_{\min}$ is close to the center of each distribution.  So while $\beta_{\text{min}}$ does not seem to be a strict limit on the field strength, it is generally gives a good description of the saturated state field strength in the stratified case in addition to the unstratified case.  

Along similar lines, \cite{simon17} have recently characterized non-uniform magnetic structure by examining the vertical magnetic field in a series of shearing box simulations.  They found that the MRI can lead to zonal flows, consistent with previous studies (see, e.g., \citealt{bai14}) which in turn create long-lived regions of concentrated vertical flux.  We observe similar features in our simulations with $Am=1.0$ and $Am=3.2$ (without FUV layers).  For more diffusive simulations ($Am<0.5$), we also observe filament-like structures of concentrated $B_z$.  However, these structures are weak and fleeting in nature when compared to those described above.  For less diffusive simulations ($Am>5$), we do not observe any significant vertical flux concentration.  This non-uniform structure to the magnetic field is likely to have an effect on the mean level of fluid turbulence within the box. However, a more detailed characterization of how the fluid turbulence is tied to these structures is beyond the scope of our paper.

\subsection{Turbulence and Stress}
Throughout this paper we will consider perturbations to various quantities, defined (for a quantity $a$) as
\begin{equation}
\delta a(x,y,z,t) \equiv a(x,y,z,t)-\langle a \rangle_{xy}(z,t).                  
\label{eqPerts}
\end{equation}
\noindent
where the $xy$ subscript on the angled bracket denotes a horizontal average. 
For vector quantities, the perturbation of each component is calculated and then the total perturbation is the sum of the 3 components in quadrature. 

\begin{figure}[h!]
\centering
\includegraphics[width=0.9\columnwidth]{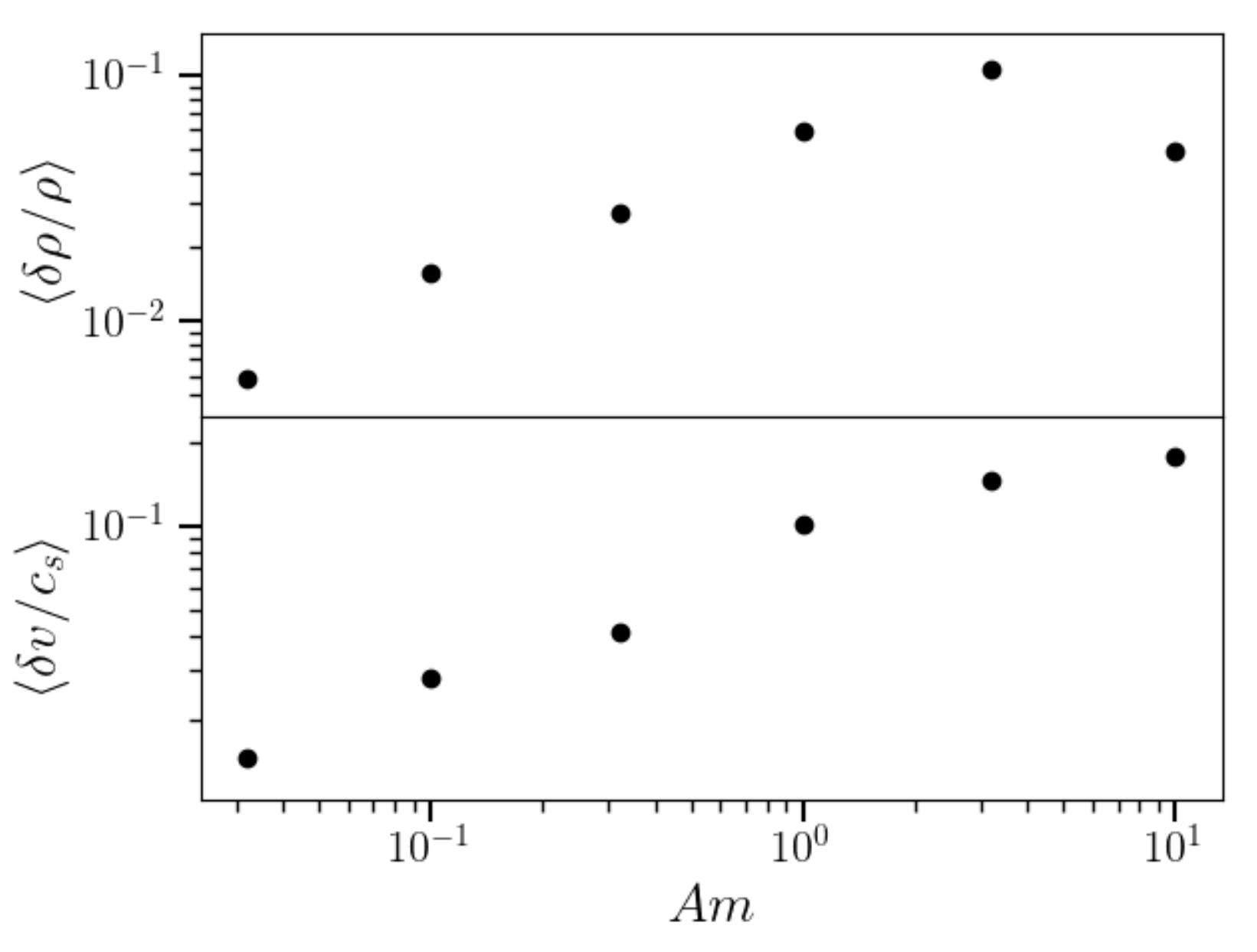}
\caption{Hydrodynamic diagnostics of turbulence $\delta \rho$ and $\delta v$ time and space averaged (within $0.5H$ of the mid-plane) for simulations SLN\_Am0.032 through SLN\_Am10.  The level of turbulence increases as the damping decreases, consistent with expectations.}
\label{figHydroPertsNFUV}
\end{figure}

The first set of turbulence diagnostics that we calculate are the perturbed $\rho$ and $v$.  In figure \ref{figHydroPertsNFUV}, we show that both of these perturbations scale as a power law with $Am$.  The strength of turbulent velocity perturbations (relative to the gas sound speed) in particular has been of much interest recently as it provides a direct comparison to observations of turbulent line widths \citep{flaherty15,flaherty17,flaherty18}. Many of these observations probe optically thick gas at large $|z|$ (usually $|z|\sim 2$--3$H$) in the outer regions of protoplanetary disk.  To date, these observations have constrained turbulence to be quite weak ($\delta v/c_{\rm s} < 0.05$) for two systems: HD163296 \citep{flaherty15,flaherty17} and TW Hya \citep{flaherty18}.  In an attempt to explain this weak turbulence, \cite{simon17} proposes a scenario in which the outer disk is shielded from ionizing sources and thus strong ambipolar diffusion, analogous to what we are simulating here. 

\begin{figure}[h!]
\centering
\includegraphics[width=0.9\columnwidth]{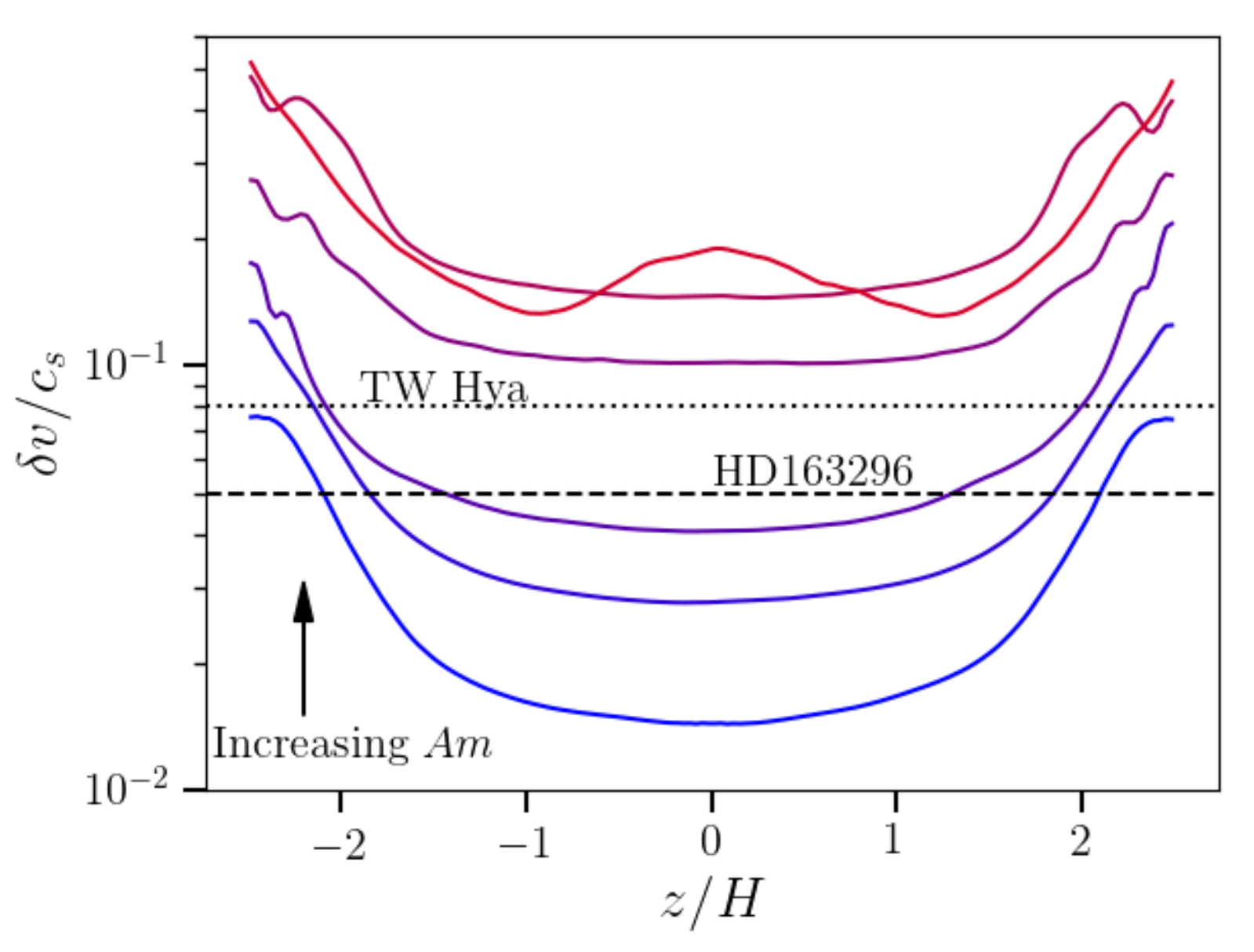}
\caption{The velocity perturbation profiles for our series of simulations with no FUV layer (constant $Am$).  The profiles are shown in a gradient from blue to red, corresponding to $Am=0.032,0.1,0.32,1.0,3.2$, and $10.0$.  The observed limits on TW Hya and HD163296 \citep{flaherty15,flaherty17,flaherty18} are shown in the dotted and dashed lines respectively. (Note that in the case of TW Hya, this limit only applies to $|z| \sim 2$--3$H$.) Even in a disk with no FUV-ionized layers, for $\beta_0=10^4$ it would take a diffusivity of at least $Am=0.032$, and likely higher, to explain the observations of weak turbulence. }
\label{figDvProfilesNFUV}
\end{figure}

Given these considerations, we plot the the vertical profiles for the perturbed velocity in Fig.~\ref{figDvProfilesNFUV}, finding (in agreement with \citealt{simon17}) that at large heights above the mid-plane, $\delta v/c_{\rm s}$
approaches values above the observational constraint.  Even in our most diffusive simulation ($Am=0.032$), turbulence at large $|z|$ is above that of HD163296 and approaches the limit from TW Hya.

While these simulations do not extend into the corona far enough to cover the entire observed region, we expect that $\delta v$ would continue to increase as $z$ increases.  More specifically, we speculate that, in the $Am=0.032$ simulation, $\delta v$  would exceed or equal the limit for TW Hya if we were to consider the region between $z/H=2$ and $3$.  So even in the scenario where a disk has no ionized layers, an even lower ionization (i.e., ambipolar diffusion; $Am < 0.032$) or a weaker background vertical field may be required to be consistent with observations. 

\begin{figure}[h!]
\centering
\includegraphics[width=0.9\columnwidth]{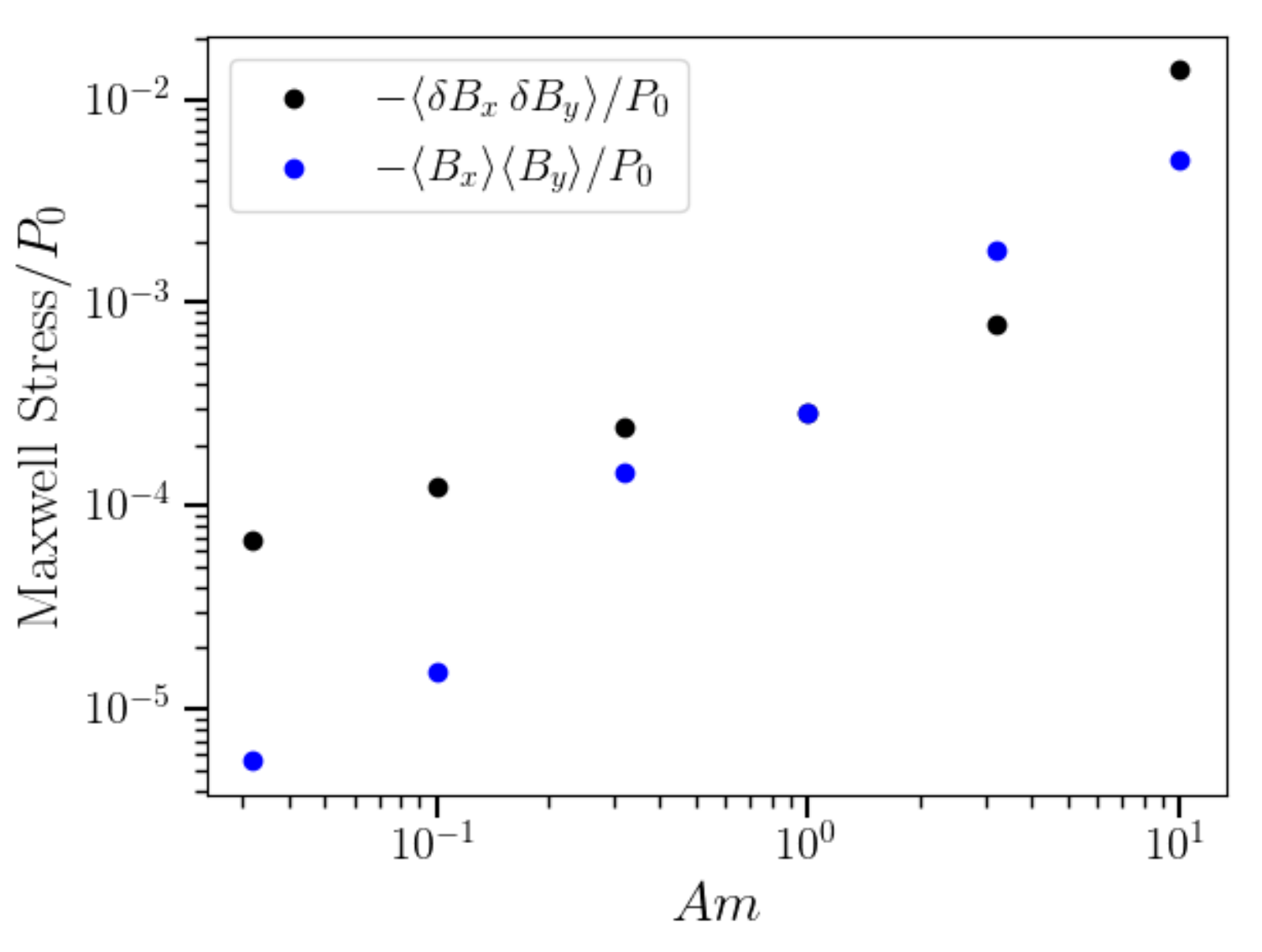}
\caption{The magnetic stress calculated from perturbations (black) and from laminar fields (blue).  Note that at $Am=1$, the points are directly on top of one another.  The stress clearly scales with $Am$.  For $Am < 1$, the stress is dominated by small-scale perturbations. }
\label{figMaxwellLsSsNFUV}
\end{figure}

Finally, we examine the Maxwell stres, and we calculate this stress in two different ways: background-subtracted field perturbations, $-\langle\delta B_x \delta B_y\rangle$, and the average field components, $-\langle B_x\rangle \langle B_y\rangle$, evaluated for each constant-$z$ plane.  These perturbation and laminar contributions are plotted in figure \ref{figMaxwellLsSsNFUV}.  The stress scales as a power law with $Am$, with perturbations providing the dominant component for $Am<1$.

\section{Stratified Ambipolar Simulations with FUV}
Having established a baseline of simulations, and accompanying diagnostics, with no FUV ionization, in this section, we analyze simulations with FUV ionization (SLF\_Am0.01 through SLF\_Am100) and consider the behaviour of the gas and magnetic field at the mid-plane as $Am_{\text{mid}}$ is varied.  We define the mid-plane, above-transition, and corona regions to be $|z/H|=$ 0 to 0.5, 1.0 to 1.5, and 3 to 3.5 respectively.  The simulations generally saturate in terms of turbulence and field strength within about 10 orbits, so the time-averaged quantities shown in the plots in this and following sections are taken starting at 10 orbits and extending to the end of the run.  This gives us a baseline of 40 orbits at minimum and over 100 orbits for most runs (see table \ref{tabSims}).

\subsection{Mid-plane field strength}
\begin{figure}[h!]
\centering
\includegraphics[width=0.9\columnwidth]{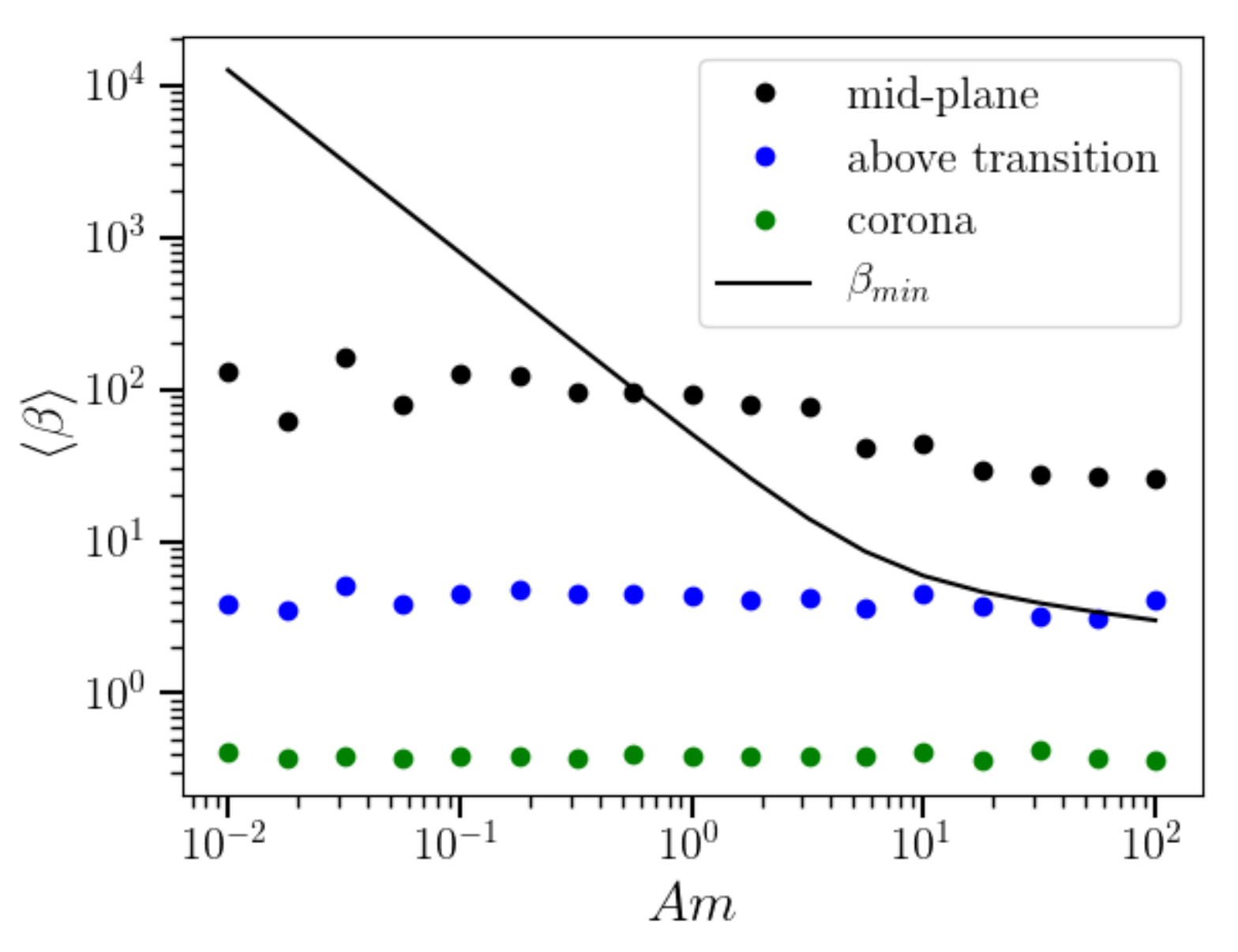}
\caption{The time-averaged $\beta$ parameter plotted versus $Am$ for SLF\_Am0.01\_S0.1 through SLF\_Am100\_S0.1 for 3 spatial regions: the mid-plane, above the transition, and the corona.  The total field strength is dominated by the toroidal component.}  The black line shows the minimum $\beta$ for the mid-plane MRI (given in equation \ref{eqMinBeta}).  Because the field strength exceeds what would be expected from the local MRI for $Am<1$, we expect that this field is sourced elsewhere and transported to the mid-plane.
\label{figBetaVsAm}
\end{figure}

We calculate the time and space averaged $\beta$ in various regions of the box for our simulations with an FUV ionized active layer.  Note that $\beta$ will be dominated by the toroidal component, as we have found that $B_y$ is the dominant component of the field. We also find that the mid-plane $\beta$  is less than the expected minimum described in the previous section for simulations with $Am_{\text{mid}} < 1$ (Figure \ref{figBetaVsAm}).   

\begin{figure}
\centering
\includegraphics[width=0.98\columnwidth]{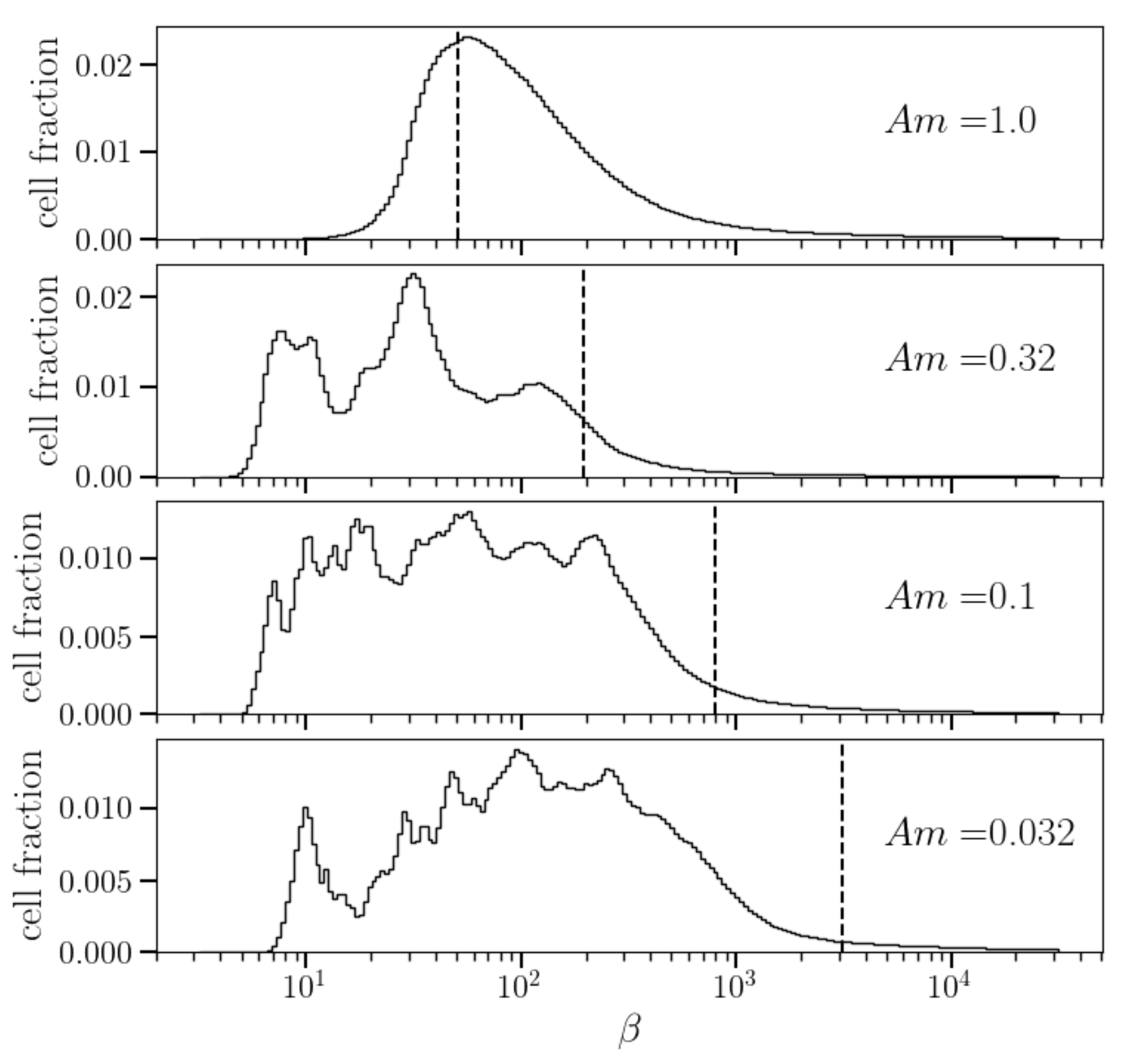}
\caption{Distribution of $\beta$ taken over the mid-plane region of SLF\_Am0.032 through SLF\_Am1.0, with the mid-plane $Am$ shown on each plot.  The vertical dashed lines indicate the minimum beta that permits the MRI for each $Am$ parameter.  While the MRI might occasionally be active in small patches, it is unlikely to dominate the dynamics of the mid-plane when $Am<1$. }
\label{figHistogram}
\end{figure}

As described briefly in section \ref{secNoFuv}, it is possible to have a situation where the average magnetic field exceeds the maximum that permits the AD-dominated MRI while still having patches of weaker field where the MRI is active.   This idea has been seen recently in similar simulations by \cite{simon17}.  Thus, simply calculating the average $\beta$ parameter is not sufficient, and so we also calculate the distribution of mid-plane $\beta$, integrated over time and for $|z|<0.5H$.  Examples of this distribution are shown in figure \ref{figHistogram}.  We find that the fraction of cells for which $\beta>\beta_{\text{min}}$ is on the order of a few percent for $Am<1$.  So while there may be small patches for which the MRI is permitted, this likely does not dominate the dynamics of the mid-plane.  We also see that the distribution for the $Am=1$ case is qualitatively different than the $Am<1$ cases, and that the $Am<1$ cases are similar to one another.  

Comparing these mid-plane field strength distributions to our simulations with no FUV layers, we see that the presence of the FUV layers widens the distributions significantly for cases with $Am<1$. In this same regime, the distributions do not tend to follow $\beta_{\text{min}}$ as a function of Am, but instead seem to approach a similar distribution regardless of the strength of the diffusion.  Lastly, at least for our fiducial choice of parameters, we do not observe any regions of concentrated vertical flux (such as those seen in \citealt{simon17}) in simulations with FUV layers.  We speculate that this is primarily due to our choice of $\Sigma_{\text{FUV}}$=0.1 g/cm$^2$, compared to $\Sigma_{\text{FUV}}$=0.01 g/cm$^2$ in \citealt{simon17}, causing our damping region to be smaller in spatial extent.  Due to the field transport mechanisms discussed in section 5, we expect that this smaller damping region will have a stronger magnetic field due to corona-down transport of $B_y$.  This stronger field will lead to stronger ambipolar diffusion, making it more difficult to build up regions of concentrated flux before they diffuse away.  This speculation is tentatively confirmed by an FUV simulation we performed with $\Sigma_{\text{FUV}}$=0.01 g/cm$^2$, a domain size of $4H$ x $8H$ x $8H$, and $Am_{\text{mid}}=0.1$.  This simulation does show regions of vertical flux concentration, although they are not as long-lived as those seen by \citealt{simon17}.

These results have two implications: 
\begin{enumerate}
\item{The magnetic field at the mid-plane is not produced by the local MRI, as the field strength exceeds what the local, AD-dominated MRI would be able to maintain for the given $Am$ parameter.  The field here must be transported from the MRI active regions above and below the mid-plane.}
\item{In the majority of the mid-plane region ($>90 \%$), the MRI cannot be active because the field strength makes ambipolar diffusion strong enough to damp field perturbations faster than they can grow. }
\end{enumerate}

\subsection{Hydrodynamic Turbulent Properties}
\begin{figure}[h!]
\centering
\includegraphics[width=0.9\columnwidth]{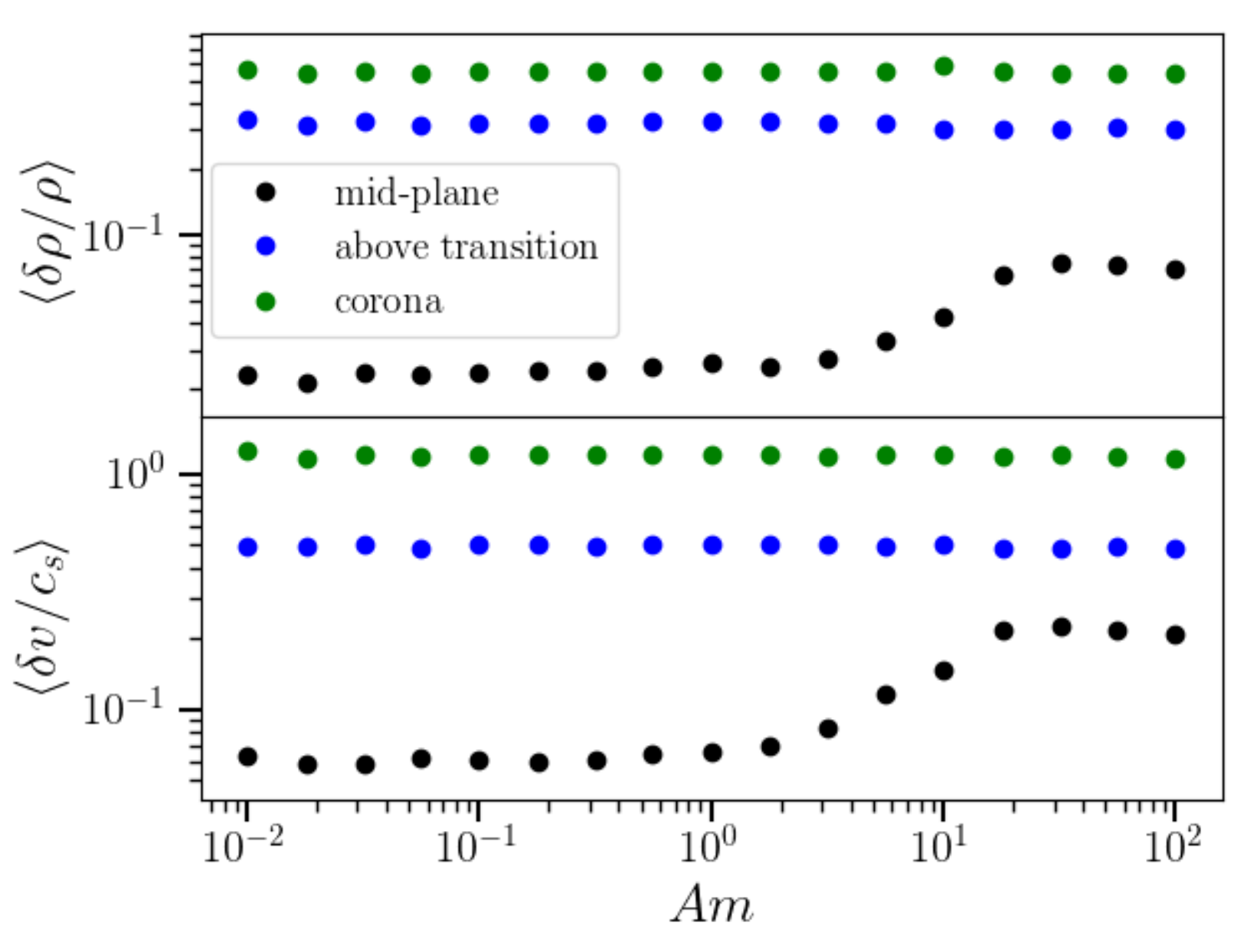}
\caption{Velocity and density perturbations for SLF\_Am0.01 through SLF\_Am100. The mid-plane quantities do not scale with $Am$ for $Am<1$ indicating that these perturbations are not generated by local MRI. In addition, the mid-plane turbulence does not seem to affect the level of turbulence in the corona.}
\label{figBothPerts}
\end{figure}

We again calculate the time and space averaged perturbations of density and velocity; in figure \ref{figBothPerts} we see that hydrodynamic diagnostics of turbulence are independent of the magnetic diffusivity of the mid-plane for $Am<1$.  This agrees with the expectation from the previous section, as the field is too strong for the MRI to be operating locally and generating density and velocity perturbations.  Any turbulence at the mid-plane has trickled down from the corona \cite[possibly related to the mechanisms seen in][for Ohmic dead zones]{fleming03,oishi09,gole16}, and given that the properties of the corona are not varied in these simulations, one would expect to recover a constant perturbation amplitude at the mid-plane as a function of $Am$ for $Am<1$.  For $Am>1$ we see a power law emerge, similar to the case with no FUV layer where we see a power law across the entire range of $Am$ tested.  This is indicative that the MRI is operating in this regime and gets stronger as $Am$ increases, as expected.  We should also note that the properties of the corona are not effected by the diffusivity of the mid-plane.  

\subsection{Stress and Magnetic field perturbations}
\begin{figure}[h!]
\centering
\includegraphics[width=0.9\columnwidth]{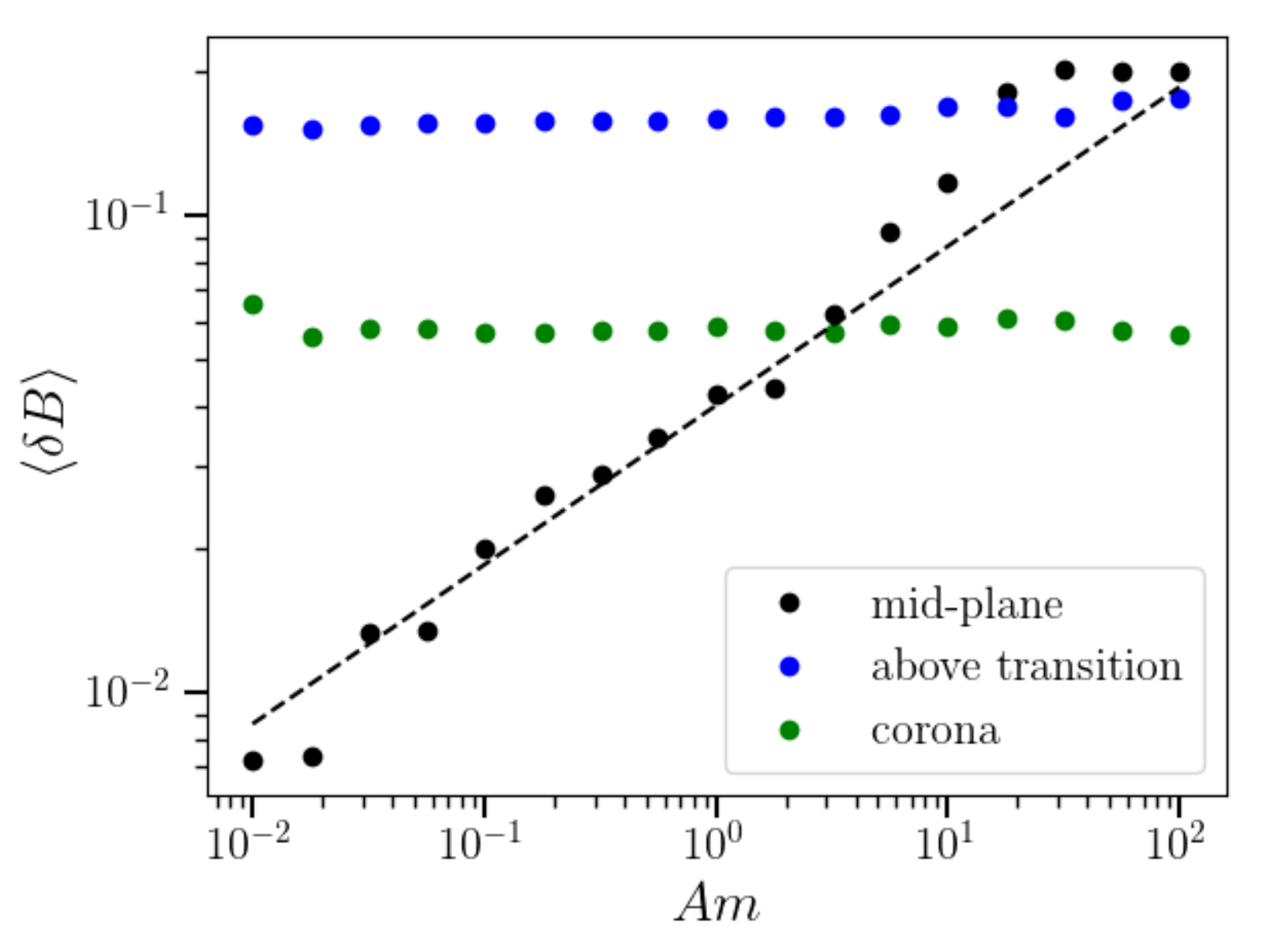}
\caption{Magnetic field perturbations versus $Am$ within different vertical regions.  The mid-plane perturbations scale roughly as $Am^{1/3}$, indicated by the black dashed line.  Despite the mid-plane scaling, the mid-plane does not influence the other regions. }
\label{figdB}
\end{figure}

In Fig.~\ref{figdB}, we calculate the average field perturbation as a function of $Am$ for different vertical regions.  The mid-plane field perturbations scale roughly as a power law of $Am$ with an index of $1/3$, even for $Am<1$.  This is in contrast to the hydrodynamic diagnostics of turbulence discussed in the previous section that did not depend on $Am$ in this regime.  Considering a simple diffusion equation for the magnetic field (e.g., equation 8), field perturbations will be diffused away in direct proportion to the magnetic diffusivity.  Thus, we see that the field perturbations scale with $Am$.  Hydrodynamic perturbations, on the other hand, have no explicit diffusion, but can be damped out by turbulent motions (e.g., through a turbulent cascade to the grid scale, where numerical diffusion dominates); thus, these particular quantities do not scale with $Am$ for $Am<1$.  

The expected $\alpha$ parameter in the ambipolar region determined by \cite{bai11a} is given by
\begin{equation}
\alpha = 1/(2\langle \beta \rangle).
\label{eqAlpha}
\end{equation}  
If we assume $\beta$ will saturate to $\beta_{\text{min}}$, we can combine this relation with equation \ref{eqMinBeta} to get an expected relationship between $Am$ and the stress for the AD-dominated MRI.  Note that this is an upper limit on this stress, as $\beta$ will not always reach $\beta_{\text{min}}$, depending on the field geometry (\citealt{bai11a} figure 15).  We do not necessarily expect to recover this relation for $Am<1$ because the MRI is not active, but will use it as a point of comparison. 

\begin{figure}[h!]
\centering
\includegraphics[width=0.9\columnwidth]{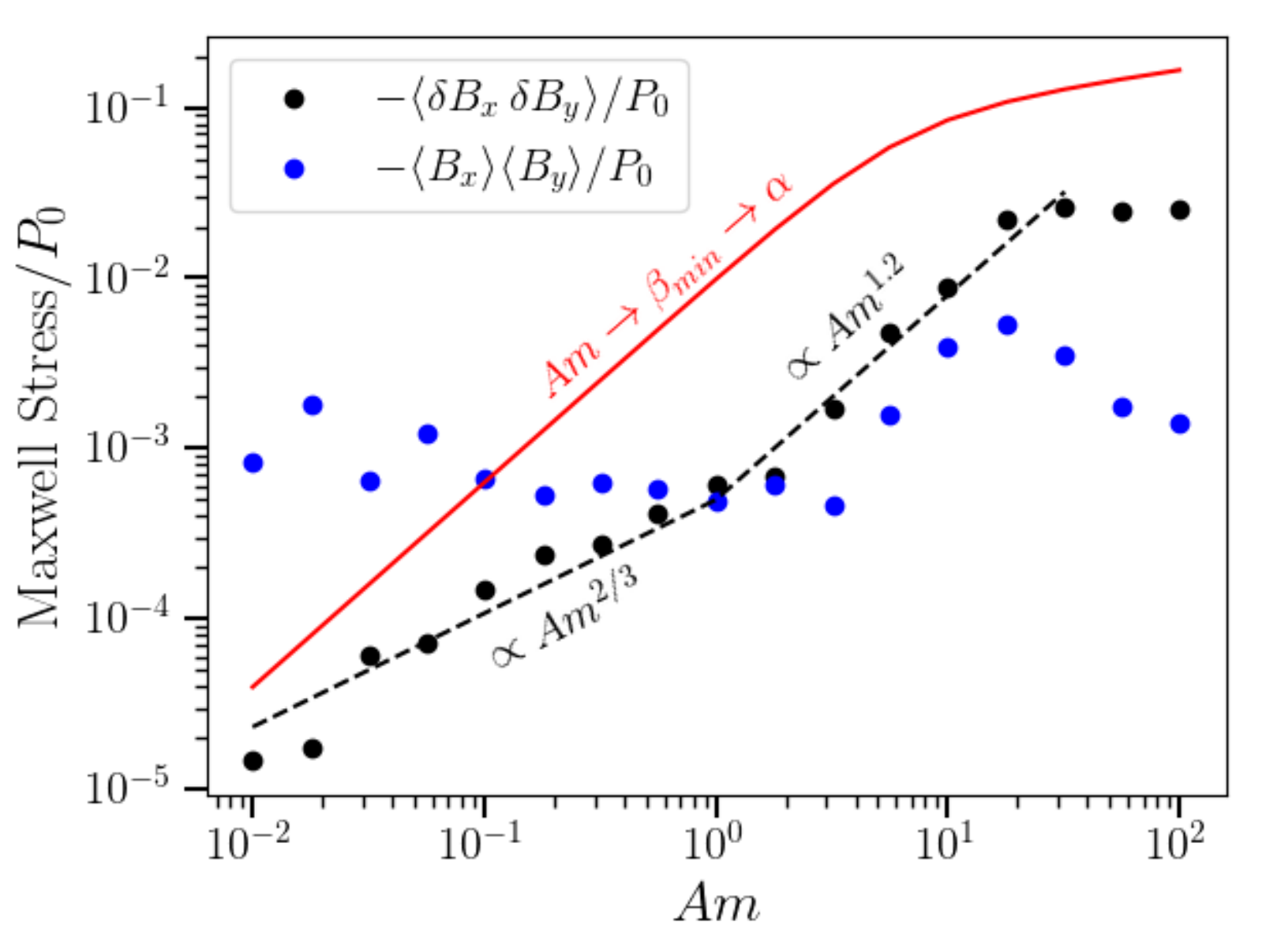}
\caption{The magnetic stress calculated from perturbations (black) and from laminar fields (blue).  The red line indicates the predicted upper limit of stress from \cite{bai11a}.  The black dashed line indicates the expected power law scaling of the stress from perturbations: $Am^{2/3}$ for $Am<1$ (based on the scaling seen in figure \ref{figdB}), and $Am^{1.2}$ for $1 \sim Am$ (based on equations \ref{eqMinBeta} and \ref{eqAlpha}).  The mid-plane stress is dominated by the laminar component for very diffusive mid-planes ($Am<1$).}
\label{figMaxwellLsSs}
\end{figure}

We again calculate the Maxwell stress from both laminar field and field perturbations, shown for the mid-plane region in figure \ref{figMaxwellLsSs}.  Given that each individual field component perturbation scales as $Am^{1/3}$, we expect and do see that the stress due to perturbations scales roughly as $Am^{2/3}$, being the multiplication of two field components.  We find that the laminar stress dominates for highly diffusive mid-planes, equaling the perturbed stress around $Am=1$.  This is qualitatively different from the results of our simulations with no FUV layers (figure \ref{figMaxwellLsSsNFUV}), where we see that the stress from magnetic field perturbations remains dominant for all simulations with $Am<1$.  

For $Am<1$, there are two effects that make the observed mid-plane stress different than the prediction in equation \ref{eqAlpha} :
\begin{enumerate}
\item As shown in figure \ref{figBetaVsAm}, $\langle B_y \rangle$ is stronger than the prediction based on $\beta_{\text{min}}$.  This will enhance the laminar magnetic stress defined to be -$\langle B_x \rangle \langle B_y \rangle$.  Figure \ref{figMaxwellLsSs} shows that this is the dominant effect on the total stress for $Am \lesssim 0.1$, where the laminar field is able to produce a stress of $\alpha \sim 10^{-3}$.  
\item The MRI is not active and is not generating perturbations locally.  Any perturbations will just be due to trickle down turbulence from the corona.  This causes the stress due to perturbations to be less than the prediction.  This is the dominant effect on the total stress for $0.1\lesssim Am \lesssim 1$, where the total stress is significantly less than the upper limit.  
\end{enumerate}

For $1 \lesssim Am$, we expect $\beta$ to scale roughly as $Am^{-1.2}$ until $Am>>1$, where the MRI will saturate and is essentially in the ideal MHD limit (equation \ref{eqMinBeta}).  Based on equation \ref{eqAlpha}, this means the stress should scale as $Am^{1.2}$.  We recover this scaling between $Am=1$ and $\sim 10$, at which point the stress saturates, remaining constant as $Am$ increases. 

We should note that, for completeness, we performed tests at higher resolution and with a larger box size in the horizontal directions to ensure that these would not affect our results (see table \ref{tabSims}).  In all of these tests we find that the behaviour is consistent with our fiducial choice of resolution and box size.


\section{Vertical transport of magnetic field in the FUV simulations}
\begin{figure}[h!]
\centering
\includegraphics[width=1.0\columnwidth]{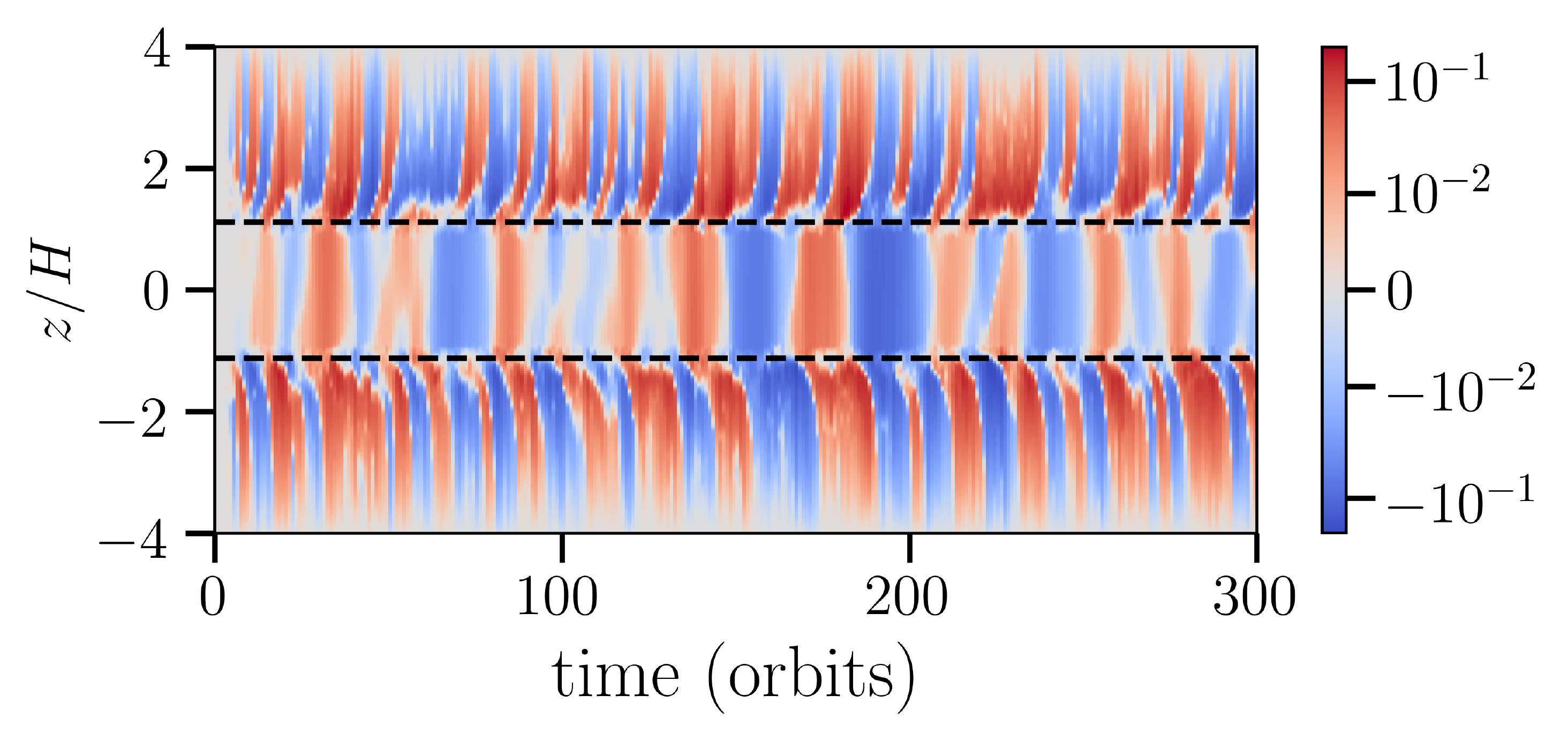}
\caption{Space-time diagram of $\text{sign}(B_y) B_y^2 / 2P_0$ for SLF\_Am0.1.  Here, $P_0$ is the mid-plane pressure.  The horizontal dashed lines indicate $z_{\text{FUV}}$. }
\label{figStBy}
\end{figure}

Figure~\ref{figStBy} shows the space-time diagram of the horizontally averaged toroidal field $B_y$ for run SLF\_Am0.1.  Upon visual inspection of the diagram, it appears that even for a AD damped mid-plane, the MRI dynamo starts at the mid-plane and then is transported upwards via buoyancy.  If this is true, it is inconsistent with our claim in the previous section that the MRI-active corona is able to transport flux to the mid-plane. In this section, we will perform a temporal correlation analysis to further elucidate this mystery.     

We compute the temporal correlation of the mean $B_y$ for each x-y plane with that at the mid-plane:   
\begin{equation}
c(z,t') = \int_{t=0}^{T} \big \langle B_{y} \big \rangle_{x,y}(z=0,t) \ast \big \langle B_{y} \big \rangle_{x,y}(z,t-t'){\rm d}t,
\label{eqCorrelation}
\end{equation}
After computing this correlation function, the time-shift ($t'$) at which $c$ is maximized is found for each $z$.  In our definition, $t' < 0$ at a given $z$ corresponds to magnetic field at that $z$ lagging that of the mid-plane. Thus, in the example of buoyantly rising magnetic fields, $t' < 0$ and $|t'|$ would increase as the field rose further away from the mid-plane.  

\begin{figure}[h!]
\centering
\includegraphics[width=0.9\columnwidth]{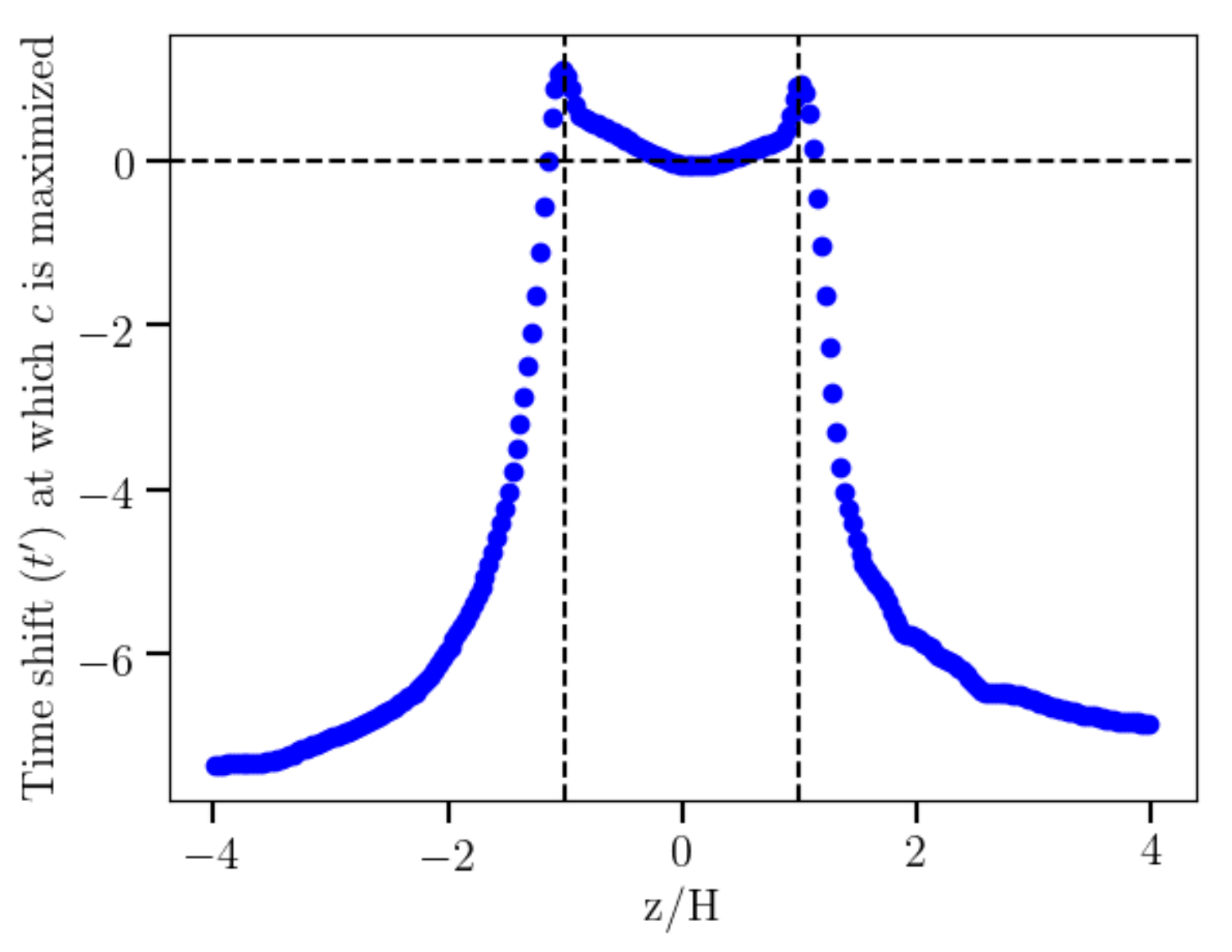}
\caption{Time shift of maximum correlation relative to the mid-plane as a function of $z$ for the SLF\_Am0.1 simulation.  The dashed vertical lines indicate the center of the transition between FUV ionization and no FUV ionization.  We see that the mid-plane lags behind the transition region, implying that diffusion towards the mid-plane is able to beat out the effect of buoyancy.}
\label{figLagVsZ}
\end{figure}

As shown in Fig.~\ref{figLagVsZ}, while field well above the transition region (denoted by the vertical dashed lines) has $t' < 0$, suggesting that field is buoyantly rising away from the transition region, the mid-plane region has $t' > 0$ suggesting that the mid-plane itself lags behind the transition region.  This result holds for all of our simulations with $Am < 10$.  The dominant region for controlling the MRI dynamo is right at the damping zone transition.  The butterfly diagram (figure \ref{figStBy}) still generally has its typical appearance because the corona still lags behind both the transition region and the mid-plane.  However, upon close inspection it can be seen that the pattern at the mid-plane shows a ``concave" shape on several occasions (specifically at orbits 80, 215, and 255).  This is consistent with ``corona-down" transport and the concave shape of the mid-plane in the temporal correlation plot (figure \ref{figLagVsZ}).
 
These correlation results are somewhat surprising, as it is generally difficult to transport magnetic flux \textit{towards} the mid-plane of a disk.  Usually the dominant effect in vertical transport of flux is magnetic buoyancy, which gives space-time diagrams like figure \ref{figStBy} their characteristic ``butterfly" appearance.  Considering a magnetic flux tube with field strength $B_{\text{pert}}$ in a disk, the magnetic flux provides additional pressure such that the gas pressure required to maintain hydrostatic balance against the vertical component of the central's stars gravity is reduced by $B_{\text{pert}}^2$ (in our units).  In the isothermal case, the gas density will also be reduced within the flux tube by $B_{\text{pert}}^2/c_s^2$.  The flux tube is then lighter than it's surroundings and experiences a force away from the mid-plane.  This effect is known as magnetic buoyancy.  The buoyant force $F_b$ on the gas in the flux tube is equal to the weight of the displaced fluid:
\begin{equation}
F_{\text{b}} = \frac{B_{\text{pert}}^2}{c_s^2} g,
\label{eqBuoyancy1}
\end{equation}
where $g$ is the acceleration due to vertical gravity.  Dividing by the gas density in the tube and using $g=\Omega^2 z$ yields a simple expression for the buoyancy acceleration $a_b$:
\begin{equation}
a_{\text{b}} = \frac{B_{\text{pert}}^2}{\rho c_s^2} \Omega^2 z =  \frac{\Omega^2 z}{\beta_{\text{pert}}},
\label{eqBuoyancy2}
\end{equation}
where $\beta_{\text{pert}}$ is the $\beta$ parameter corresponding to $B_{\text{pert}}$.  In order to arrive at a buoyant time-scale $t_b$, we consider the time it would take for a perturbation of constant $\beta$ to rise $H$ at this rate of acceleration.  This yields
\begin{equation}
t_b = \Omega^{-1}\bigg(\frac{2H}{z} \beta_{\text{pert}} \bigg)^{1/2}.
\label{eqBuoyancyTime}
\end{equation}

While the flux tube is accelerating upwards it will also be diffusing due to ambipolar diffusion.  In order to transport a significant portion of the flux to the mid-plane, this diffusive effect will have to occur on roughly the same time-scale as buoyancy or shorter.  The ambipolar diffusion time under some simplifying assumptions is given in equation \ref{eqTad}.  Over length-scale $H$, the diffusion time is $t_{\text{AD}} = 2 \beta_{\text{pert}} Am\Omega^{-1}$.  Taking the ratio of these two time-scales yields:
\begin{equation}
\frac{t_d}{t_b} = \bigg( \frac{z}{h} \bigg)^{1/2}\beta_{\text{pert}}^{1/2}\hspace{1mm} Am.
\end{equation}
Note that $h=H/\sqrt{2}$.  In other words, diffusion will be the dominant transport mechanism if $Am < Am_{\text{crit}}$, where $Am_{\text{crit}} = \big( \frac{z}{h} \beta_{\text{pert}} \big)^{-1/2}$.  If we consider an MRI-generated perturbation of $\beta=30$ at $z/H=1$, $Am_\text{crit} \sim 0.18$.  This is roughly consistent with the results of our simulations with FUV layers, specifically figure \ref{figHistogram}, which shows that the mid-plane magnetic field strength distribution changes in nature somewhere between $Am=0.1$ and $1$.  This likely corresponds to a transition from diffusion dominated transport to buoyancy dominated transport.

If $\Sigma_{\text{FUV}}$ is decreased, perturbations will start at larger $|z/H|$ and we expect the effect of the corona on the mid-plane will lessen due to two effects:  
\begin{enumerate}
\item The mid-plane is farther from the initial perturbation, and the time-scale for diffusive transport scales as length squared.
\item Buoyant acceleration is $\propto z/H$, so perturbations will rise faster from their starting position (this is accounted for in the relations given above).
\end{enumerate}

We should also note the potential effect that our choice of vertical seed field ($\beta_z=10^4$) might have on these results regarding vertical field transport from ambipolar diffusion.  A weaker vertical field will generate a weaker MRI and less turbulence, as seen in the non-ideal MHD simulations by \cite{simon17}.  This will subsequently lessen the strength of the toroidal field perturbations of the saturated-state MRI.  Weaker perturbations will both diffuse more slowly and buoyantly rise more slowly (equations \ref{eqTad} and \ref{eqBuoyancyTime} respectively).  However, Equation 18 shows that $t_d/t_b \propto \beta_{\text{pert}}^{1/2}$; thus, the relative strength of buoyancy will increase compared to ambipolar diffusion in the case of a weaker net-field.  



\section{Conclusions}

In order to further understand the properties of turbulence in the outer regions of protoplanetary disks, we have run a series of stratified shearing box simulations with varying strength of ambipolar diffusion and with and without an FUV layer.  Our main conclusions are as follows:

\begin{enumerate}
\item Our simulations without an FUV layer generally match the behaviour of the similar but unstratified simulations of \cite{bai11a}.  The $\beta$ parameter at the mid-plane has a narrow distribution centered near $\beta_{\text{min}}$, and both hydrodynamic and magnetic diagnostics of turbulence scale with $Am$.  This is consistent with the AD-damped MRI operating.      
\item In our simulations with an FUV layer and $Am < 1$ within the mid-plane region, we have demonstrated that the behaviour of the mid-plane, even qualitatively, does not match the behaviour of unstratified or stratified simulations without an FUV layer.  This is due to the interaction with the MRI-active layers, which can transport magnetic flux to the mid-plane to a point such that the MRI is no longer permitted.  There are several important differences compared to the no-FUV-layer case:  
\begin{itemize}
\item The magnetic stress is dominated by laminar fields rather than small scale field perturbations.  
\item Hydrodynamic perturbations do not scale with $Am$ at the mid-plane; they are set by the fluid motions in the FUV layers rather than local MRI.
\item For most of the box, $\beta<\beta{\text{min}}$.  The field strength distribution is wider than the no-FUV case and does not vary greatly as a function of $Am$. 
\end{itemize}
All of these points are consistent with the MRI being arrested in most of the mid-plane region by the magnetic field.  The dynamics of the mid-plane are primarily determined by the FUV layers instead of the local MRI.  While the MRI may operate in small patches of weak field, this does not seem to be a dominant effect and occurs in a small fraction of the box, if at all.
\item For $Am\lesssim10$, toroidal field at the mid-plane is temporally correlated with toroidal field in the corona with a lag time on the order of an orbit.  This correlation implies that diffusion (turbulent or ambipolar) of flux towards the mid-plane is able to overcome magnetic buoyancy, which is usually the dominant effect in the vertical transport of magnetic flux in a disk. 
\item The properties of the MRI-active region are not affected by the diffusivity of the mid-plane region. 
\end{enumerate}

What do these results imply for studies of turbulence in protoplanetary disks?  First and foremost, if an FUV layer is present, there is a significant influence of the strongly ionized layers on the highly diffusive mid-plane.  Thus, one cannot in general treat each vertical layer of the disk independently; analytic or semi-analytic models that attempt to do so will potentially be missing physical effects (e.g., vertical diffusion of magnetic field) that will change estimates for the degree of turbulence and angular momentum transport.  In the absence of an FUV layer, this issue is less clear however.  While there will be no interaction with the mid-plane from an FUV layer for these simulations, we speculate that buoyancy will still play an important role and that one should be cautious in assuming that each layer of the disk can be treated independently, even without an FUV layer.

Furthermore, our results (particularly the second and fourth conclusions above) are quite relevant to recent observational \citep{flaherty15,flaherty17,flaherty18} and theoretical \citep{simon17} results that suggest a picture in which turbulence in the outer regions of protoplanetary disks is quite weak. If indeed, FUV photons are somehow prevented from reaching the outer disk, as suggested by \cite{simon17}, then the turbulent velocity in the disk is largely set by the strength of the ambipolar-diffusion-dominated MRI.  Considering those simulations without FUV,  $\delta v/c_{\rm s} \sim 0.01$ within the mid-plane region for the most diffusive case and rises up to $\delta v/c_{\rm s} \sim 0.05$ at $|z| \sim 2.5H$, as shown in Fig.~\ref{figDvProfilesNFUV}. Given the general trend of increasing $\delta v$ with $|z|$, it seems likely that at larger $|z|$, the turbulent velocity will be marginally consistent with observations (which show $\delta v/c_{\rm s} \sim 0.05$--$0.1$; \citealt{flaherty17,flaherty18}) at best, though potentially still larger than observational constraints.  In either case, this result indicates that even for very diffusive disks (which can arise from small grains soaking up charges, or significant reduction of ionizing flux), to be consistent with observations, one needs to reduce the background magnetic field strength \citep{bai11a,simon17}. 

In summary, our results provide a more in-depth understanding of turbulence and angular momentum transport in the outer regions of protoplanetary disks. Not only will this aid in future modeling of these disks, but with observations of such systems continuing well into the next decade, such work is crucial to making progress in comparing theoretical expectations with observational data.

\acknowledgments
We thank Phil Armitage for general advisement, Xuening Bai for useful comments, and Colin McNally for helpful conversations.  We acknowledge support from NASA under awards NNX13AI58G, NNX14AB42G and 80NSSC18K0640 from the Origins of Solar Systems and Astrophysics Theory programs.  This work utilized the RMACC Summit supercomputer, which is supported by the National Science Foundation (awards ACI-1532235 and ACI-1532236), the University of Colorado Boulder, and Colorado State University. The Summit supercomputer is a joint effort of the University of Colorado Boulder and Colorado State University.  We also acknowledge the Texas Advanced Computing Center (TACC) at The University of Texas at Austin for providing HPC resources that have contributed to the research results reported within this paper. URL: http://www.tacc.utexas.edu.

\bibliographystyle{mnras}
\bibliography{refs}

\end{document}